\documentclass[a4paper,fleqn]{cas-dc}

\usepackage{graphicx}
\usepackage[justification=centering]{caption}
\usepackage{subcaption}
\usepackage{multirow}
\usepackage{enumitem}
\usepackage{listings}
\usepackage{xspace}
\usepackage[flushleft]{threeparttable}
\usepackage[numbers]{natbib}
\usepackage{color,soul}
\usepackage{mdframed}
\usepackage{amsmath}

\definecolor{light-gray}{gray}{0.92}  
\newenvironment{gtheorem}%
  {\begin{mdframed}[backgroundcolor=light-gray]\begin{mdtheorem}{name}{label}}%
  {\end{mdtheorem}\end{mdframed}}

\def\tsc#1{\csdef{#1}{\textsc{\lowercase{#1}}\xspace}}
\tsc{WGM}
\tsc{QE}
\tsc{EP}
\tsc{PMS}
\tsc{BEC}
\tsc{DE}

\newdefinition{definition}{Definition}

\definecolor{ao}{rgb}{0.0, 0.5, 0.0}
\definecolor{mauve}{rgb}{0.8, 0.7, 1.0}
\newcommand{\etal}{\textit{et al.}\space}
\newcommand{\tool}{\textsc{Empica}\xspace}
\newcommand{\codellama}{Code Llama\xspace}
\newcommand{\chatgpt}{GPT-3.5\xspace}
\newcommand{\magiccoder}{MagicCoder\xspace}
\newcommand{\deepseek}{DeepSeek-Coder\xspace}

\newcommand{\robust}{Robustness to \textit{SP} transform.\xspace}
\newcommand{\sensitive}{Sensitivity to \textit{SNP} transform.\xspace}

\begin{document}
\let\WriteBookmarks\relax
\def\floatpagepagefraction{1}
\def\textpagefraction{.001}

\shorttitle{\tool}

\shortauthors{Nguyen \textit{et~al.}}

\title [mode = title]{An Empirical Study on Capability of Large Language Models in Understanding Code Semantics}    


\author{Thu-Trang Nguyen}[orcid=0000-0002-3596-2352]
\ead{trang.nguyen@vnu.edu.vn}

\affiliation{organization={Faculty of Information Technology, VNU University of Engineering and Technology},
    city={Hanoi},
    country={Vietnam}}

\author{Thanh Trong Vu}
[orcid=0009-0008-3377-6565]
\ead{thanhvu@vnu.edu.vn}

\author{Hieu Dinh Vo}[orcid=0000-0002-9407-1971]
\ead{hieuvd@vnu.edu.vn}

\author{Son Nguyen}[orcid=0000-0002-8970-9870]
\ead{sonnguyen@vnu.edu.vn}
\cormark[1]

\cortext[cor1]{Corresponding author}

\maketitle

\begin{abstract}
Large Language Models for Code (code LLMs) have demonstrated remarkable performance across various software engineering (SE) tasks, increasing the application of code LLMs in software development. 
Despite the success of code LLMs, there remain significant concerns about the actual capabilities and reliability of these models, \textit{``whether these models really learn the semantics of code from the training data and leverage the learned knowledge to perform the SE tasks''}. 
In this paper, we introduce \tool, a comprehensive framework designed to systematically and empirically evaluate the capabilities of code LLMs in understanding code semantics. Specifically, \tool systematically introduces controlled modifications/transformations into the input code and examines the models' responses. 
Generally, code LLMs must be \textit{robust to semantically equivalent code inputs} and be \textit{sensitive to non-equivalent ones} for all SE tasks.
Specifically, for every SE task, given an input code snippet $c$ and its semantic equivalent variants, code LLMs must robustly produce consistent/equivalent outputs while they are expected to generate different outputs for $c$ and its semantic non-equivalent variants.
Our experimental results on three representative code understanding tasks, including code summarization, method name prediction, and output prediction, reveal that the robustness and sensitivity of the state-of-the-art code LLMs to code transformations vary significantly across tasks and transformation operators. In addition, the code LLMs exhibit better robustness to the semantic preserving transformations than their sensitivity to the semantic non-preserving transformations. These results highlight a need to enhance the model's capabilities of understanding code semantics, especially the sensitivity property.

\end{abstract}

\section{Introduction}

Large Language Models (LLMs) for code have demonstrated remarkable performance on a variety of automated programming tasks. Automating such tasks is inherently complex, requiring the model to understand numerous concepts within the underlying code. However, despite this effectiveness on various code-related tasks, a fundamental question remains: \textit{How deeply do these large language models for code (code LLMs) truly understand the code they interact with?}
Current evaluations usually assess the usefulness (e.g., $pass@k$ metric) of programs generated by these models based on the passed tests for the given tasks in several benchmarks such as HumanEval~\cite{HumanEval}, MBPP~\cite{MBPP}, and CodeContests~\cite{CodeContests}. While these task-driven evaluations measure end-to-end performance, they fail to reveal the LLMs' capabilities in understanding code semantics, which could be crucial for ensuring the reliability of code LLMs. 

Prior studies~\cite{icse22-syntax,blackbox,ma2023scope} have begun to explore the capabilities of LLMs in code understanding. However, a deeper understanding of the LLMs' capabilities to capture semantic aspects, including control dependence and data dependence, remains elusive. 
Wan~\etal~\cite{icse22-syntax} prioritize syntactic correctness over semantic understanding. Evaluation methods might assess if generated code compiles without truly grasping its intended functionality.
Hooda~\etal~\cite{blackbox} evaluate LLMs' capability in understanding code semantics through a single task, which is code completion. This restricts our understanding of how broadly code LLMs can understand code.
%
%
Ma~\etal~\cite{ma2023scope} evaluate the code LLMs' capabilities in reconstructing code syntax and semantic structures by fine-tuning the LLMs to predict certain properties in code such as nodes and edges in Abstract Syntax Tree (AST) or data/control dependence. It is unclear how LLMs arrive at their outputs, making it difficult to assess their reasoning process and identify potential biases in semantic understanding.

In this paper, we propose \tool, a novel framework for \textbf{systematically and empirically evaluating code LLMs' capabilities in understanding essential code semantic aspects}.
Our approach involves analyzing the \textit{sensitivity} and \textit{robustness} of code LLMs to changes in code semantics. 
Within this framework, we focus on two fundamental semantic relations in the programming analysis: \textit{control dependence} and \textit{data dependence}~\cite{pdg1,pdg2,pdg3}.
%
%
To assess code LLMs' capabilities in understanding these dependencies, we define eight code transformations. These transformations can be categorized as either \textit{semantic non-preserving} (affecting the semantic relations) or \textit{semantic preserving} (not affecting the semantic relations).

We apply our framework to three critical software engineering tasks with varying levels of semantic complexity: code summarization, method name prediction, and output prediction.
We evaluate the models for each task without fine-tuning or additional training data to minimize potential biases. 
%
%
%
Ideally, models must recognize semantic changes introduced by non-preserving transformations and adjust their outputs accordingly.
Conversely, their outputs must remain consistent for programs before and after semantic-preserving transformations.
Our work incorporates four state-of-the-art (SOTA) code LLMs: \deepseek~\cite{deepseek-coder}, \codellama~\cite{codellma}, \magiccoder~\cite{magicoder}, and \chatgpt~\cite{gpt}. 
To reduce potential out-of-distribution threats, we analyze how these models perform on the code generated by these models themselves when subjected to the aforementioned semantic code transformations. 
This analysis will provide valuable insights into the strengths and weaknesses of current SOTA models in terms of code semantic comprehension.

Our experimental results demonstrate substantial variations in the robustness and sensitivity of code LLMs to the code transformations applied. These variations are observed across different SE tasks and transformation operators. Notably, the models exhibit a greater degree of robustness to semantic-preserving transformations compared to their sensitivity towards semantic non-preserving transformations. These results highlight the necessity for further development in code LLMs' capabilities, particularly in enhancing their sensitivity to changes in code semantics. 

The contributions of this paper are as follows:

\begin{enumerate}
    \item Evaluation Framework: We introduce \tool, a framework for systematically evaluating LLM understanding of essential semantic aspects in code.
    \item Empirical Results: We present an empirical evaluation using \tool on various tasks and benchmark datasets.
    \item Findings: We analyze the results to identify the strengths and weaknesses of current SOTA code LLMs regarding semantic code comprehension.
\end{enumerate}

The detailed experimental results and source code for reproducing experiments can be found at \textit{\url{https://ttrangnguyen.github.io/EMPICA/}}.

\textbf{Paper Organization}. We present the background and related studies in Section~\ref{sec:background} and describe the motivation for our
framework in Section~\ref{sec:motivation}. Our proposed framework, \tool, is described in Section~\ref{sec:framework}. Section~\ref{sec:eval} and \ref{sec:results} present the evaluation setup and experimental results. Finally, Section~\ref{sec:conclusion} concludes our work.
\section{Background and Related Work}
\label{sec:background}


\subsection{Large Language Models for Code}

Large Language Models for Code (code LLMs) are becoming popular in the Software Engineering (SE) industry. Multiple code LLMs, such as Github Copilot\footnote{https://github.com/features/copilot}, AlphaCode~\cite{li2022competition}, and \chatgpt~\cite{gpt}, have been employed to speed up the software development process. These models play a significantly important role in the automation of multiple SE tasks such as code completion~\cite{zhang2023repocoder}, program repair~\cite{jiang2023impact}, or test generation~\cite{schafer2023empirical}, etc. As the sizes of these models and the amount of training data have increased, code LLMs have exhibited impressive performance in these tasks~\cite{zhang2023repocoder, jiang2023impact, schafer2023empirical}.

Existing LLMs for code are constructed based on the Transformer~\cite{transformer} (encoder-decoder or decoder-only). These models are trained on extensive, unlabelled corpora of code-related data, enabling them to generate code in various programming languages. The training process typically involves the causal language modeling objective, where the model learns to predict the next token in a sequence. The success of LLMs in the SE task can be attributed to three main factors: Large Size, Premium Data, and Expert Tuning~\cite{nl2code}. 

Recent advancements in code LLMs demonstrate a trend towards larger model sizes, accompanied by superior performance~\cite{xu2022systematic, liu2024your}. These studies show that increasing the number of model parameters can significantly enhance model capabilities. For instance, early models such as CodeT5~\cite{codet5} had limited success due to their relatively small size. In contrast, modern models like Codex, Code Llama, and DeepSeek achieve remarkable performance on various benchmarks with their massive number of parameters.

Additionally, high-quality code corpora enable LLMs to learn different programming languages and coding paradigms effectively. To ensure the quality of the training corpus, it is common for LLMs to perform extensive data preprocessing on the large amount of data collected. One common strategy is removing likely auto-generated or unfinished code files. Additionally, specific rules are employed to filter out uncommon code files. These rules include the repository star rating, file size, line length, and alphanumeric rate.

Expert tuning is also essential for optimizing model performance. Common practices include using the Adam optimizer or its variants and further training from the models pre-trained on text. Key hyperparameters such as learning rate, batch size, window size, warmup steps, gradient accumulation steps, and sampling temperature require expert tuning. Specialized tokenizers, such as Byte-level Byte-Pair-Encoding and SentencePiece, are applied on code corpora to handle programming syntax effectively.

\subsection{Evaluating Code LLMs}

Several benchmarks have been established to evaluate the capabilities of code LLMs in various SE tasks. For example, Chen~\etal~\cite{HumanEval} introduced \textit{HumanEval}, a popular benchmark designed to assess the functional correctness of the generated code in multiple programming languages. Similarly, Austin~\etal~\cite{MBPP} proposed \textit{MBPP} dataset, which contains solvable problems for entry-level programmers. This dataset is also widely used to evaluate the program synthesis of code LLMs~\cite{touvron2023llama, chowdhery2023palm, shinn2024reflexion}. Another manually crafted benchmark \textit{ClassEval}~\cite{du2023classeval} focuses on evaluating LLMs' performance on class-level code generation. Experimental results on these benchmarks have demonstrated the remarkable performance of code LLMs across various SE tasks, which significantly accelerates the software development process.

In addition, many approaches~\cite{liu2024reliability, liu2024your, blackbox} proposed to further thoroughly test code LLMs in various code generation and code understanding tasks. Liu \etal~\cite{liu2024reliability} conducted comprehensive experiments on eight models using five benchmarks to examine the capabilities and limitations of these models. In another research, Lie \etal~\cite{liu2024your} evaluates whether the code generated by code LLMs is really correct by introducing more sophisticated test cases for the problems in the benchmark \textit{HumanEval}. Their experimental results show that the performance of code LLMs at \textit{pass@k} for the problems in \textit{HumanEval} significantly decreased when additional test cases were included. Besides, Hooda~\etal~\cite{blackbox} assessed the code completion capabilities of code LLMs by perturbing concept predicates such as control flow or data flow in the input programs. Their evaluations of ten LLMs reveal that these perturbations frequently led to incorrect code completion by the models.

Furthermore, the abilities of code LLMs in understanding the other aspects of the programs have also been investigated~\cite{chen2024evaluating, karmakar2023inspect, maunveiling}. Chen \etal~\cite{chen2024evaluating} evaluated how code LLMs capture the runtime behavior of program execution such as code coverage or execution path prediction. Karmakar \etal~\cite{karmakar2023inspect} and Ma \etal~\cite{maunveiling} proposed probing tasks for assessing the code LLMs capability in capturing code syntax and semantics aspects, including Abstract Syntax Tree (AST), Control Flow Graph (CFG), and Control Dependence Graph (CDG). Their results indicate that while models can effectively capture syntactic relationships, they still struggle to understand semantic relationships.
\section{Motivation}
\label{sec:motivation}


Despite the success of code LLM in SE tasks, there are multiple doubts about the performance of these models~\cite{blackbox, ma2023scope, yang2024unveiling}, ``\textit{Whether code LLM really learn the syntax and semantics of code from the training data and leverage the learned knowledge to conduct SE tasks}''. 
Yang \etal~\cite{yang2024unveiling} have conducted experiments to investigate the memorization of LLMs for code with CodeParrot and CodeParrot-small. Their experimental results show that from 43\% to 57\%, outputs of these models contain memorized information. This raises a question about the reliability of the answers provided by code LLM.

\begin{figure}
   \begin{subfigure}{1\columnwidth}
\centering
\begin{lstlisting}[language=Java]
class Problem {
    public static boolean belowZero(ArrayList<Long> operations) {
        long balance = 0;
        for (long operation : operations) {
            balance += operation;
            if (balance < 0) {
                return true;
            }
        }
        return false;
    }
}
\end{lstlisting}
\caption{Example of a code snippet}
\label{fig:motivation_code_correct}
\end{subfigure}

\begin{subfigure}{1\columnwidth}
\centering
 \includegraphics[width=1\columnwidth]{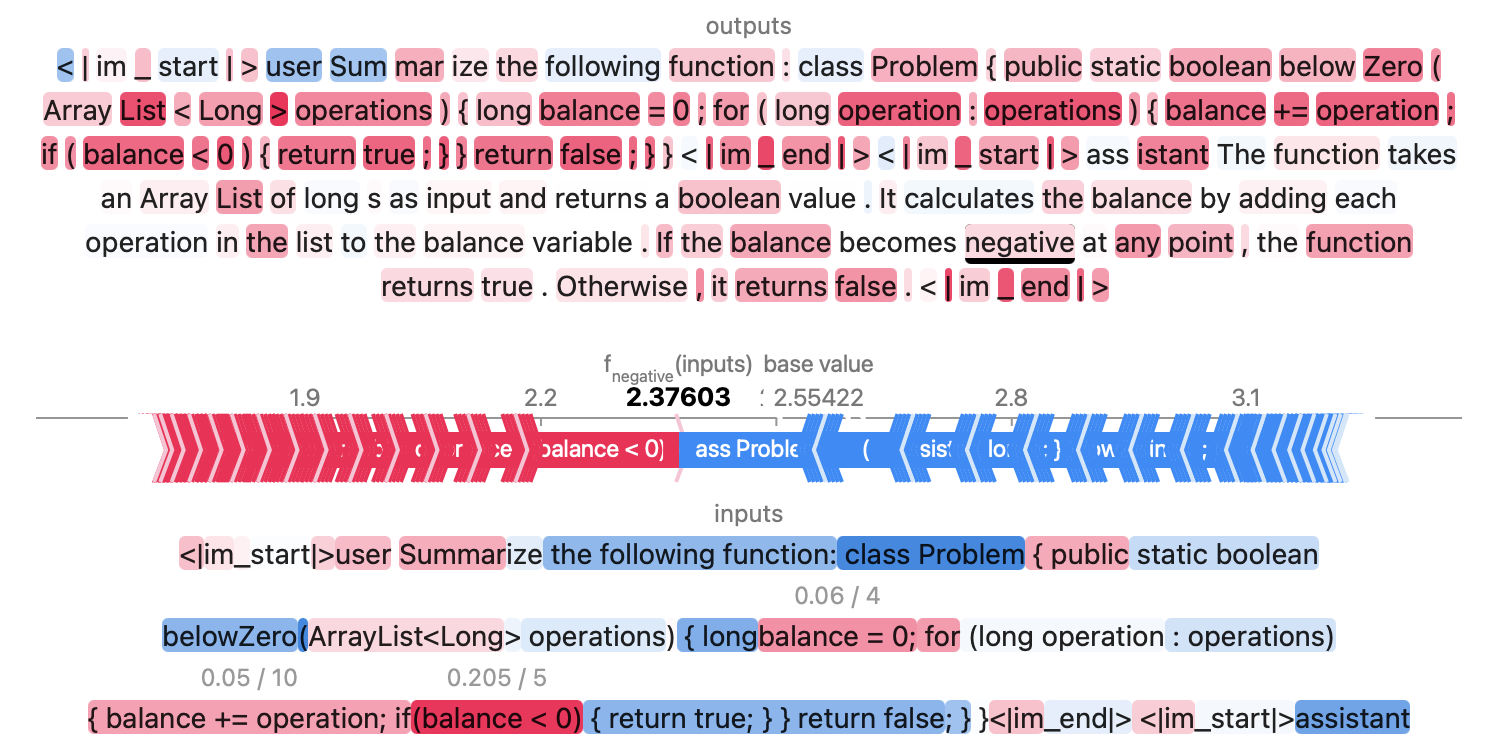}
\caption{The explanation for the summary of the code in Figure~\ref{fig:motivation_code_correct}}
\label{fig:motivation_explanation_correct}
\end{subfigure}
\caption{An code summary generated by Microsoft/phi-2 and the corresponding explanation generated by SHAP}
\label{fig:motivation_example_correct}
\end{figure}

\begin{figure}
   \begin{subfigure}{1\columnwidth}
\centering
\begin{lstlisting}[language=Java]
class Problem {
    public static boolean belowZero(ArrayList<Long> operations) {
        long balance = 0;
        for (long operation : operations) {
            balance += operation;
            return true;
        }
        return false;
    }
}
\end{lstlisting}
\caption{Example of a code snippet}
\label{fig:motivation_code}
\end{subfigure}

\begin{subfigure}{1\columnwidth}
\centering
 \includegraphics[width=1\columnwidth]{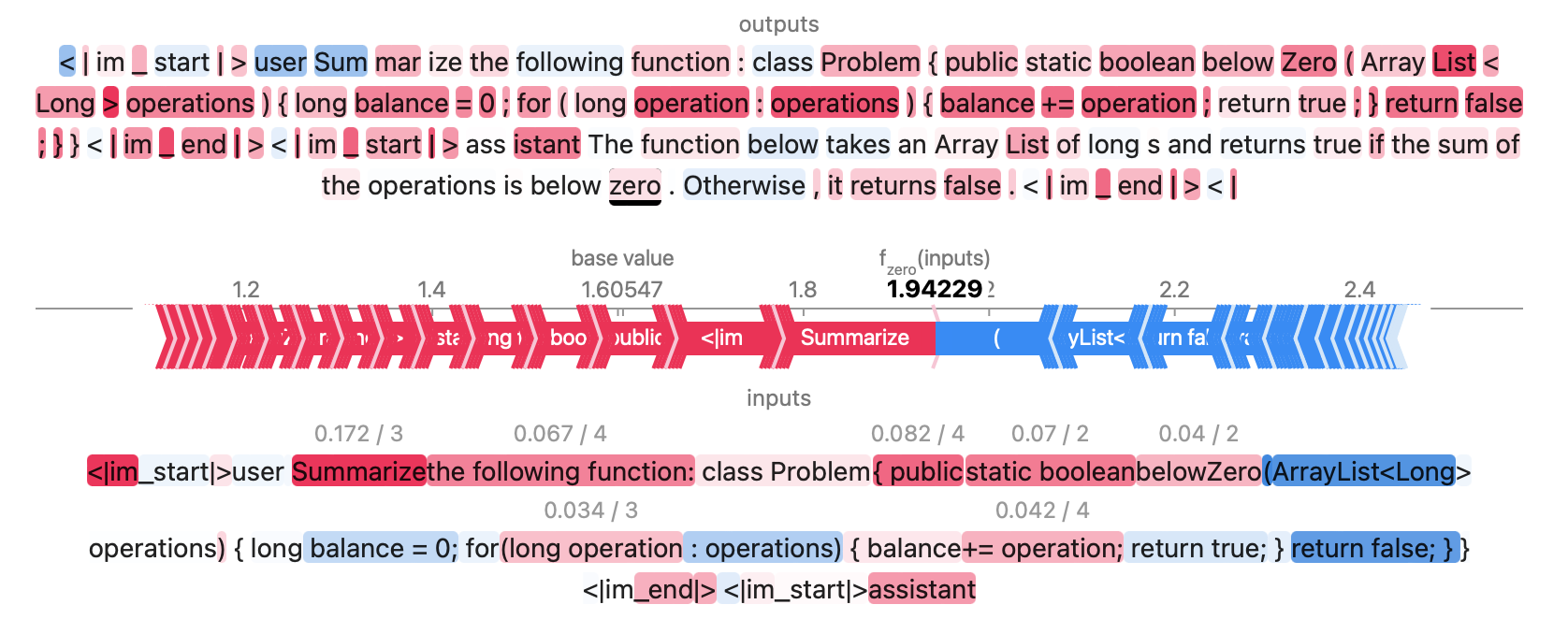}
\caption{The explanation for the summary of the code in Figure~\ref{fig:motivation_code}}
\label{fig:motivation_explanation}
\end{subfigure}
\caption{An code summary generated by Microsoft/phi-2 and the corresponding explanation generated by SHAP}
\label{fig:motivation_example}
\end{figure}

Figure~\ref{fig:motivation_example_correct} shows a summary generated by Microsoft/phi-2\footnote{https://huggingface.co/microsoft/phi-2} for a given code snippet.
This given code detects and returns \texttt{true} if, at any point, the account balance falls below zero. The generated summary is \textit{``If the balance becomes negative at any point, the function returns true.''} (Figure~\ref{fig:motivation_explanation_correct}). According to the explanation of SHAP~\cite{shap}, the model can accurately produce this summary statement, especially the token \textit{``negative''}, because it focuses on the conditional statement at line $6$, \texttt{if (balance < 0)}.

In addition, without checking the negative value of balance (e.g. \texttt{if (balance < 0)}),  the code snippet in Figure~\ref{fig:motivation_code} is semantically different from one in Figure~\ref{fig:motivation_code_correct}.
However, the summary (Figure~\ref{fig:motivation_explanation}) generated by Microsoft/phi-2 is still very similar to the summary in Figure~\ref{fig:motivation_explanation_correct}. 
Specifically, 
the summary for the code in Figure~\ref{fig:motivation_code} is  ``\textit{The function below takes an ArrayList of longs and \textbf{returns true if the sum of the operations is below zero}. Otherwise, it returns false}'' (Figure~\ref{fig:motivation_explanation}). 
As seen, this summary is incorrect regarding the given method (Figure~\ref{fig:motivation_code}).  

The explanation produced by SHAP (Figure~\ref{fig:motivation_explanation}) shows that Microsoft/phi-2 mainly focused on several tokens like \texttt{boolean}, \texttt{operation}, \texttt{true}, \texttt{false}, or especially the method name \texttt{belowZero} to generate the summary.
For example, the model focused on the expression \texttt{balance += operation} and \texttt{return true;}  in the input for producing token \textit{``if''} in the summary. Also, to generate token \textit{``zero''} in the summary, the model focused on the tokens like \texttt{boolean}, \texttt{belowZero}, and \texttt{operation} in the code snippet.
This example demonstrates that LLM could heavily rely on general patterns of input code to generate summaries without really comprehending them.

The above analysis raises the question on the capability of code LLM in understanding code semantics: ``\textit{Whether and how the code LLM can \textbf{really} understand code semantics or just capture and rely on the general patterns from the data for conducting downstream tasks}''. In this research, we propose \tool, a framework for systematically evaluating the capability of code LLM in capturing code semantics, including control and data dependencies.

\section{\tool: A Framework for Evaluating Code Semantic Understandability of  LLMs}
\label{sec:framework}

\begin{figure*}
    \centering
    \includegraphics[width=1\textwidth]{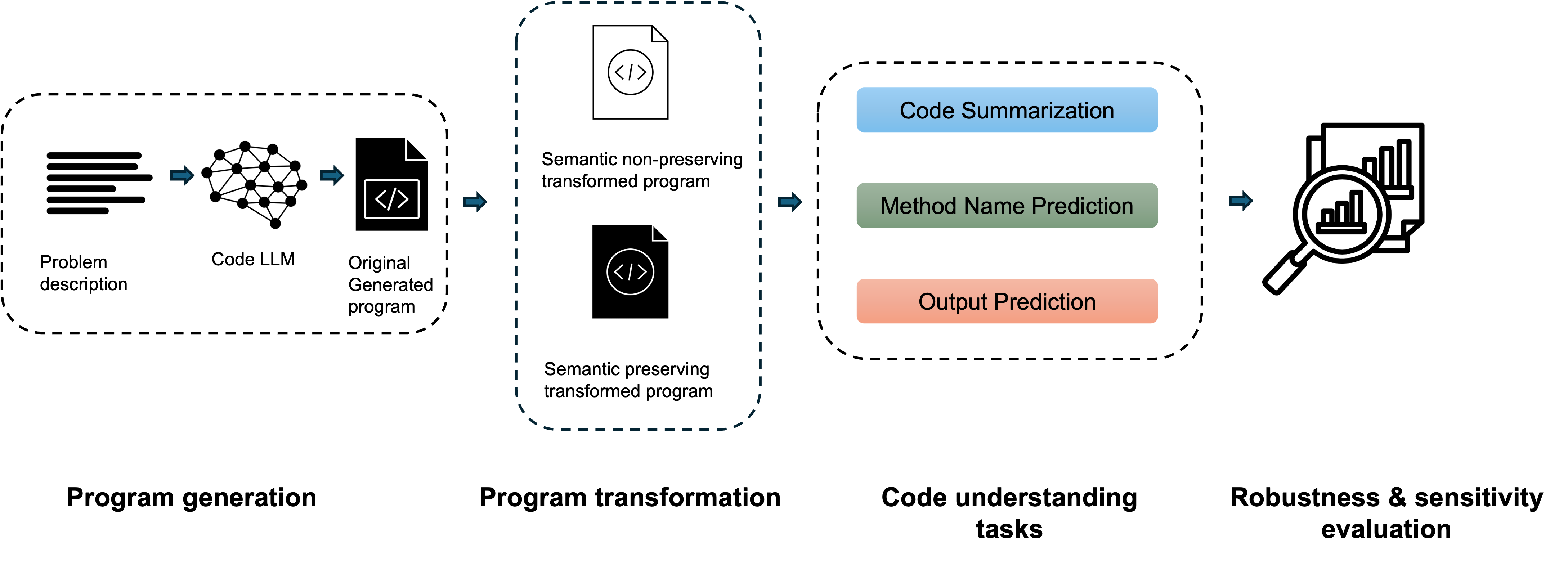}
    \caption{Framework overview}
    \label{fig:framework_overview}
\end{figure*}

This work aims to evaluate code LLMs' capability in code semantic understanding in the two fundamental code semantic relations: \textit{control dependence} and \textit{data dependence}.
Figure~\ref{fig:framework_overview} illustrates the overview of our framework, \tool, containing four main components: \textit{(1) Program Generation, (2) Program Transformation, and (3) Code Understanding, and (4) Robustness \& Sensitivity Evaluation}. 

First, for a code LLM $\mathscr{M}$ under evaluation, \tool employs $\mathscr{M}$ to generate code. Then, \tool directly uses the generated code to evaluate the capabilities of $\mathscr{M}$ in understanding code semantics.
%
%
%
This is reasonable because evaluating $\mathscr{M}$ in understanding arbitrary code could hinder the capability of $\mathscr{M}$. Indeed, $\mathscr{M}$ could misunderstand an entirely new code snippet, especially if the code significantly deviates from the patterns and distributions seen during training (i.e., \textit{out-of-distribution})~\cite{wheat-chaff}. Therefore, \tool conducts the analysis on the code generated by the model itself for a more impartial evaluation of the model's capability and to mitigate the potential risk of capability hindering due to the unseen data.

Next, \tool systematically introduces controlled modifications/transformations into the input code and observes how the model reacts in $\mathscr{M}$'s output for a specific code understanding task. 
%
\tool applies a variety of transformations designed to target both control and data dependencies in the input code snippets. We employ both \textit{semantic-preserving transformations} and \textit{semantic-non-preserving transformations} to modify the code snippets. 
Specifically, given a code snippet, \textit{semantic-preserving transformations} create various variants of the code while maintaining the original functionality of the snippet in its variants. In contrast, \textit{semantic-non-preserving transformations} produce variants whose functionality differs from the original ones.

Ideally, if $\mathscr{M}$ is capable of understanding code semantics, $\mathscr{M}$' predictions must be \textit{robust to the semantic-preserving (\textit{SP}) transformations} and \textit{sensitive to the semantic-non-preserving (\textit{SNP}) ones}.
For a specific task, $\mathscr{M}$ must produce the equivalent responses to the original input code and its variants created by the \textit{SP} transformations. In contrast, $\mathscr{M}$'s outputs for the original input and its variants modified by the \textit{SNP} transformations are expected to be different.

For thorough evaluation, \tool selects three representative SE tasks that heavily require code understanding, including \textit{code summarization}, \textit{method/function name prediction}, and \textit{output prediction}. These tasks are in different complexity levels, encompassing both static and dynamic behaviors, as well as varying degrees of program understanding, ranging from coarse-grained to fine-grained. \textit{Code summarization} requires an understanding of the overall purpose of the code snippet to provide a concise summary. \textit{Method/Function name prediction} demands exact comprehension of code semantics and functionality to predict an appropriate name for a method/function. \textit{Output prediction} involves precisely capturing the underlying logic of code and the relationship between input-output to provide correct prediction. By incorporating these tasks in different natures and complexities, \tool enables a comprehensive assessment of code LLMs in capturing code semantics.

\subsection{Semantic-aware Code Transformations}

Given a code snippet $c \in \mathcal{C}$, a \textit{transformation operator} $t$: $\mathcal{C} \to \mathcal{C}$ is a function that maps $c$ from and to a space $\mathcal{C}$ of all possible code snippets in the given programming language. Let's assume we have an oracle function $\gamma$: $\mathcal{C} \to \mathcal{S}$, which maps from the space of code snippets $\mathcal{C}$ to the specification space $\mathcal{S}$ which represents the code behaviors. A code transformation $t$ can be categorized into two groups based on its impact on $\gamma(t(c))$, semantic-preserving (\textit{SP}) and semantic-non-preserving (\textit{SNP}) transformations.

\begin{definition}{\textbf{Semantic-preserving (\textit{SP}) transformation}.}
Given a code snippet $c \in \mathcal{C}$, a transformation $t_p$ is a semantic-preserving transformation if the modifications introduced by applying $t_p$ on $c$ do not affect its specified behaviors, $\forall c \in \mathcal{C}, \gamma(c) \equiv \gamma(t_p(c))$.
\end{definition}

\begin{definition}{\textbf{Semantic-non-preserving (\textit{SNP}) transformation}.}
Given a code snippet $c \in \mathcal{C}$, a transformation $t_n$ is a semantic-non-preserving transformation if the modifications introduced by applying $t_n$ on $c$ change its specified behaviors, $\forall c \in \mathcal{C}, \gamma(c) \not\equiv \gamma(t_n(c))$.    
\end{definition}

\tool curated the transformation operators from the literature~\cite{le2024evaluating, li2024mutation, risse2023limits}. 
To systematically investigate the understanding of code LLMs in control and data dependencies, we manage the transformations with respect to these relationships. Table~\ref{tab:transformation_operators} introduces the details of the transformation operators and the corresponding examples.
\begin{table*}[]
\caption{Details of transformation operators and examples}
\label{tab:transformation_operators}
\begin{tabular}{l|l|l|l|l}
\toprule
\textbf{\begin{tabular}[c]{@{}l@{}}Transformation \\ group\end{tabular}}                            & \textbf{Relationship}                                                         & \textbf{\begin{tabular}[c]{@{}l@{}}Transformation\\ operator\end{tabular}} & \textbf{\begin{tabular}[c]{@{}l@{}}Original\\ code\end{tabular}}                                    & \textbf{\begin{tabular}[c]{@{}l@{}}Transformed\\ code\end{tabular}}                                \\ \midrule
\multirow{4}{*}{\begin{tabular}[c]{@{}l@{}}Semantic\\ preserving (\textit{SP})\\ transformation\end{tabular}}     & \multirow{2}{*}{\begin{tabular}[c]{@{}l@{}}Control\\ dependence\end{tabular}} & \begin{tabular}[c]{@{}l@{}}Convert\\ For/While\end{tabular}                & \begin{tabular}[c]{@{}l@{}}for (int i = 0; i \textless n; i++)\{\\ //do something\\ \}\end{tabular} & \begin{tabular}[c]{@{}l@{}}int i = 0;\\ while (i \textless n)\{\\ //do something\\ i++;\\ \}\end{tabular} \\ \cmidrule{3-5} 
                                                                                                    &                                                                               & \begin{tabular}[c]{@{}l@{}}Flip \\ IfElse\end{tabular}                     & \begin{tabular}[c]{@{}l@{}}if (condition)\{\\ //do A\\ \}else\{\\ //do B\\ \}\end{tabular}          & \begin{tabular}[c]{@{}l@{}}if (!condition)\{\\ //do B\\ \}else\{\\ //do A\\ \}\end{tabular}        \\ \cmidrule{2-5} 
                                                                                                    & \multirow{2}{*}{\begin{tabular}[c]{@{}l@{}}Data \\ dependence\end{tabular}}   & \begin{tabular}[c]{@{}l@{}}Rename \\ variables\end{tabular}                & int data = 10;                                                                                      & int VAR1 = 10;                                                                                     \\ \cmidrule{3-5} 
                                                                                                    &                                                                               & \begin{tabular}[c]{@{}l@{}}Reorder \\ parameters\end{tabular}              & \begin{tabular}[c]{@{}l@{}}void foo(int a, int b)\{\\ //do something\\ \}\end{tabular}              & \begin{tabular}[c]{@{}l@{}}void foo(int b, int a)\{\\ //do something\\ \}\end{tabular}             \\ \midrule
\multirow{4}{*}{\begin{tabular}[c]{@{}l@{}}Semantic\\ non-preserving (\textit{SNP})\\ transformation\end{tabular}} & \multirow{2}{*}{\begin{tabular}[c]{@{}l@{}}Control\\ dependence\end{tabular}} & \begin{tabular}[c]{@{}l@{}}Negate\\ relational\\ condition\end{tabular}    & \begin{tabular}[c]{@{}l@{}}if (condition)\{\\ //do something\\ \}\end{tabular}                      & \begin{tabular}[c]{@{}l@{}}if(!condition)\{\\ //do something\\ \}\end{tabular}                     \\ \cmidrule{3-5} 
                                                                                                    &                                                                               & \begin{tabular}[c]{@{}l@{}}Remove\\ conditional\\ statement\end{tabular}   & \begin{tabular}[c]{@{}l@{}}//do A\\ if (condition)\{\\ //do B\\ \}\end{tabular}            & \begin{tabular}[c]{@{}l@{}}//do A\\ //do B\end{tabular}                                            \\ \cmidrule{2-5} 
                                                                                                    & \multirow{2}{*}{\begin{tabular}[c]{@{}l@{}}Data\\ dependence\end{tabular}}    & \begin{tabular}[c]{@{}l@{}}Replace \\ arithmetic \\ operator\end{tabular}  & int result = a + b;                                                                                 & int result = a \% b;                                                                               \\ \cmidrule{3-5} 
                                                                                                    &                                                                               & \begin{tabular}[c]{@{}l@{}}Remove \\ def\\ statement\end{tabular}          & \begin{tabular}[c]{@{}l@{}}int foo()\{\\ int data = 10;\\ return data + 1;\\ \}\end{tabular}        & \begin{tabular}[c]{@{}l@{}}int foo()\{\\ return data + 1;\\ \}\end{tabular}                        \\ \bottomrule
\end{tabular}
\end{table*}

\subsection{Code Understanding: Subject Tasks}

In this work, \tool conducts Counterfactual Analysis in three representative SE tasks which require code understanding, including \textit{code summarization, method name prediction}, and \textit{output prediction}. 
The general idea is that given an input code snippet $c$, \textit{if a code LLM $\mathscr{M}$ is capable of understanding the code semantics, the outputs of $\mathscr{M}$ must be consistent and robust for $c$ and its variants produced by any SP transformations. In contrast, the outputs of $\mathscr{M}$ must be different and sensitive for $c$ and its variants produced by SNP ones.}  
In other words, $\mathscr{M}$ must produce semantically similar/identical outputs for $c$ and for any of $c$'s variants, $c_p$, produced by \textit{SP} transformations, $c_p = t_p(c)$, in any code understanding task $\mathscr{T}$. Conversely, $\mathscr{M}$ must recognize any modifications affecting the meaning of the program to produce different outputs for $c$ and any of its variants, $c_n$, produced by \textit{SNP} transformations, $c_n = t_n(c)$, in task $\mathscr{T}$.
The rationale of this hypothesis lies in the semantic equivalence between $c$ and $c_p$ and the semantic distinction between $c$ and $c_n$. If the outputs of $\mathscr{M}$ with the inputs $c$, $c_p$, and $c_n$ are not consistent and reliable in such way, it suggests that $\mathscr{M}$ is struggling with the code semantics comprehension.

For a dataset $\mathscr{D}$ containing $n$ code snippets which originally generated by code LLM $\mathscr{M}$ and a transformation operator $t$, let $c_i \in \mathscr{D}$, $1 \leq i \leq n$ and its corresponding transformed code be $c'_i = t(c_i)$. For a specific code understanding task $\mathscr{T}$, the outputs of $\mathscr{M}$ for $c_i$ and $c'_i$ is $o_i$ and $o'_i$, respectively, $\mathscr{T}(\mathscr{M}, c_i) = o_i$ and $\mathscr{T}(\mathscr{M}, c'_i) = o'_i$. 
The robustness of $\mathscr{M}$ is specified by the similarity of its outputs for $c_i$ and $c'_i$, i.e., the similarity between $o_i$ and $o'_i$. The sensitivity of $\mathscr{M}$ is the distinction of $\mathscr{M}$'s outputs for $c_i$ and $c'_i$, i.e., the difference between $o_i$ and $o'_i$. Specifically, the robustness and sensitivities of $\mathscr{M}$ for the code snippets in  $\mathscr{D}$ with the transformation operator $t$ could be calculated by the following equations.

\begin{equation}
\label{eq:robustness}
Robustness(\mathscr{M}, \mathscr{T}, t) = \frac{1}{n}\sum^n_i sim(o_i, o'_i)
\end{equation}

\begin{equation}
\begin{split}
\label{eq:sensitivity}
Sensitivity(\mathscr{M}, \mathscr{T}, t) & =  \frac{1}{n}\sum^n_i [1-sim(o_i, o'_i)] \\ 
                                    & =  1 - Robustness(\mathscr{M}, \mathscr{T}, t)
\end{split}
\end{equation}

In the Equation~\ref{eq:robustness}, $sim$ is a similarity function which measures the semantic similarity of two outputs $o_i$ and $o'_i$, $sim(o_i, o'_i) \in [0, 1]$, while the semantic distinction between $o_i$ and $o'_i$ is the complement of their similarity, $[1-sim(o_i, o'_i)] \in [0, 1]$. 
%
%
Ideally, $\mathscr{M}$'s outputs must be robust to the \textit{SP} transformations and sensitive to the \textit{SNP} ones. 
In other words, $\mathscr{M}$ is \textbf{\textit{adequate}} in understanding code semantics when $Robustness(\mathscr{M}, \mathscr{T}, t)$ is close to $1$ to any \textit{SP} transformation operator $t$. At the same time, $Sensitivity(\mathscr{M}, \mathscr{T}, t)$ is also close to $1$ to any \textit{SNP} transformation operator $t$. Otherwise, it suggests a \textbf{\textit{deficiency}} of $\mathscr{M}$ in understanding code semantics.
Note that the similarity function could differ for each code understanding task.

\subsubsection{Code Summarization}
\textit{Code summarization} ($\mathscr{T}_S$) is the task of concisely describing a code snippet's overall functionality, purpose, and behavior in a natural language~\cite{gao2023code}. To accurately provide a summary of a given code snippet, the model needs to precisely capture the semantics of the code. Thus, we employ code summarization as a representative task for evaluating code LLMs' capability in understanding code semantics.  


Given a code snippet $c_i \in \mathscr{D}$, which is originally generated by code LLM $\mathscr{M}$, and $c'_i = t(c_i)$ is the transformed version of $c_i$ which generated by applying a transformation operator $t$. If $t$ is a \textit{SP} transformation, the functionality of $c'_i$ is semantically equivalent to $c_i$, and the summary of $c'_i$ is expected to be similar/identical with that of $c_i$. Meanwhile, if $t$ is a \textit{SNP} transformation, the functionality of $c'_i$ is different from $c_i$, and their corresponding summaries are expected to be semantically different from each other.




Let $d_i$ and $d'_i$ be the summaries of the code snippets $c_i$ and $c'_i$, both outputted by $\mathscr{M}$. To measure the \textit{Robustness} and \textit{Sensitivity} of $\mathscr{M}$ in the code summarization task, we compare the similarities of the summaries $d_i$ and $d'_i$. Particularly, the robustness in the Equation~\ref{eq:robustness} is realized with the similarity function as follows: 



\begin{equation}
\label{eq:code_summarization}
sim(d_i, d'_i) = cosine(v_i, v'_i)
\end{equation}

In the equation~\ref{eq:code_summarization}, the similarity of the summaries is measured by the $cosine$ similarity.
Specifically, $cosine$ measures the similarity of $v_i$ and $v'_i$, two representation vectors of $d_i$ and $d'_i$, respectively. 
In \tool, we measure their similarities regarding both lexical and semantic aspects. For \textit{lexical similarity}, $d_i$ and $d'_i$ are represented by Term Frequency-Inverse Document Frequency (TF-IDF) vectors. For \textit{semantic similarity}, the meanings of $d_i$ and $d'_i$ are encoded by a pre-trained embedding model such as Sentence-BERT~\cite{reimers2019sentence}, which has been widely applied to derive the representation for sentences in natural language~\cite{minaee2021deep, kashyap2024comprehensive}.

\subsubsection{Method Name Prediction}
\textit{Method name prediction} ($\mathscr{T}_N$) is the task of suggesting an appropriate method/function name for a given code snippet~\cite{li2021context}. Successful method name prediction requires a deep understanding of the underlying semantics and operations of the code. Similar to code summarization, method name is another form of summary of code functionality, yet it is more succinct. 

Let $m_i$ and $m'_i$ be the method names predicted by Code LLM $\mathscr{M}$ for the code snippet $c_i$ and its transformed code $c'_i$, respectively. To measure the capability of $M$ in understanding code semantics for this task, $\mathscr{T}_N$, we compare the similarity of $m_i$ and $m'_i$ at \textit{token level} (i.e., exactly match) and \textit{sub-token level} (i.e., precision, recall, and f1-score).

For \textit{token-level} assessment, the similarity of $m_i$ and $m'_i$, $sim(m_i, m'_i)$, is measured regarding whether they are exactly identical, $sim(m_i, m'_i) = EM(m_i, m'_i)$. Specifically, $EM(m_i, m'_i)$ measures whether $m_i$ and $m'_i$ are exactly the same, $EM(m_i, m'_i) = 1$ if $m_i$ and $m'_i$ are identical, otherwise,  $EM(m_i, m'_i) = 0$.

For \textit{sub-token level} assessment, $sim(m_i, m'_i)$ is measured regarding how similar their sub-tokens are. \tool tokenizes $m_i$ and $m'_i$ by the \texttt{under\_score} and \texttt{camelCase} conventions, then compares how similar these sub-token sets are. Let $K_i = \{k_{i1}, ..., k_{ij}\}$ and $K'_i = \{k'_{i1}, ..., k'_{im}\}$ be the sub-token sets of $m_i$ and $m'_i$, respectively. The similarity of $K_i$ and $K'_i$ is measured in terms of \textit{precision}, \textit{recall}, and \textit{F1-score}. In particular, they are measured as follows:

$$precision(m_i, m'_i) = \frac{|K_i \cap K'_i|}{|K'_i|}$$ 

$$recall(m_i, m'_i) = \frac{|K_i \cap K'_i|}{|K_i|}$$

$$F1(m_i, m'_i) = 2 \times \frac{precision(m_i, m'_i) \times recall(m_i, m'_i)}{precision(m_i, m'_i) + recall(m_i, m'_i)}$$ 
%
%
In summary, to evaluate the code semantics understandability of $\mathscr{M}$ in this task, $\mathscr{T}_N$, in Equation~\ref{eq:robustness}, the similarity function, $sim(o_i, o'_i)$ could be expressed by $EM(m_i, m'_i)$, $precision(m_i, m'_i)$, $recall(m_i, m'_i)$, or $F1(m_i, m'_i)$.




\subsubsection{Output Prediction}

The task of \textit{output prediction}, ($\mathscr{T}_O$), involves predicting the output of a given code snippet when it is executed with a specific input~\cite{chen2024evaluating}. 
%
%
To achieve high performance in this task, code LLMs must comprehensively understand the code's semantics, operation logic, and execution flow. 
For \textit{code summarization} and \textit{method name prediction}, these tasks are only required to capture the program's \textit{static} behaviors. Meanwhile, output prediction needs to analyze not only \textit{static} but also \textit{dynamic} behaviors of the given code snippets. Thus, the correct prediction of the test output is more complex than the previous two.

Let $k$ be the number of test inputs of $c_i$ and its transformed code $c'_i$. Let
$o_{ij}$ and $o'_{ij}$, $1 \leq j \leq k$ be the outputs of $c_i$ and $c'_i$ which predicted by code LLM $\mathscr{M}$ with the $j^{th}$ test. If $c'_i$ is created by a \textit{SP} transformation, the behaviors and functionality of $c_i$ and $c'_i$ are equivalent. They behave similarly with the same input, i.e., return equal outputs for a given input. On the other hand, if $c'_i$ is created by a \textit{SNP} transformation, the functionalities of $c_i$ and $c'_i$ are different. They could produce different outputs for a given input.
Specifically, 
to assess the robustness and sensitivity of $\mathscr{M}$ in the output prediction task, $\mathscr{T}_O$, Equation~\ref{eq:robustness} is realized with the similarity function expressed as follows: 


\begin{equation}
\label{eq:output_prediction}
sim(o_i, o'_i) =
  \begin{cases}
    1       & \quad \text{if } o_i = o'_i\\
    0  & \quad \text{otherwise}
  \end{cases}
\end{equation}


\subsection{Prompt Design}

In this work, \tool utilizes prompting to evaluate code LLMs with the code understanding tasks. 
The design of the prompt can significantly impact the performance of the models. Thus, we aim to design simple prompts to mitigate the risk that the models could perform poorly because of misunderstanding the prompts.
To design appropriate prompts for each task, we evaluated and optimized the prompts via some trial queries.
Note that each LLM could have different prompt formats; we follow their original papers and the provided templates on \textit{HuggingFace} to adapt for prompt design. 
Generally, the prompt template is defined as follows:

``\texttt{Given the following code snippet [CODE]. [TASK DESCRIPTION].}''

\noindent where \texttt{[CODE]} specifies the input code and \texttt{[TASK DESCRIPTION]} specifies the query for the task. For example, for code summarization, the \texttt{[TASK DESCRIPTION]} is ``\texttt{Please summarize the given code snippet}''. For method name prediction, the \texttt{[TASK DESCRIPTION]} is ``\texttt{Please generate the method name for the given code snippet}''. For output prediction, the \texttt{[TASK DESCRIPTION]} is ``\texttt{Please complete the following test case: [TEST]}'', e.g., \texttt{assert fibfib(2) == <FILL>}.

\section{Evaluation Design}
\label{sec:eval}
To evaluate code LLMs' capability in understanding code semantics, we seek to answer the following questions:
\begin{itemize}
   
    \item \textbf{RQ1: \textit{Robustness and Sensitivity Analysis}.} How robust are the state-of-the-art code LLMs~\cite{codellma,deepseek-coder,gpt,magicoder} to the \textit{SP} transformations in code understanding tasks? How sensitive are code LLMs to \textit{SNP} transformations in these tasks?

    \item \textbf{RQ2: \textit{Model Size Analysis}.} Does code LLMs' size affect their capability to understand code semantics?
    
    \item \textbf{RQ3: \textit{Correctness Correlation Analysis}.} Is there any correlation between the code semantic understandability of code LLMs and the correctness of their generated code?

    \item \textbf{RQ4: \textit{Semantic Inference Analysis}.} Can the code LLMs correctly capture the code semantics and directly infer the relations including control and data dependence?
  
\end{itemize}

\subsection{Experimental Procedure}
To mitigate the risk of the out-of-distribution problem~\cite{berend2020cats}, \tool utilizes the code generated by the model under evaluation to evaluate the model's capability of understanding code semantics. In this work, we employ the popular benchmark \textit{HumanEval}~\cite{HumanEval}, which is widely used in code generation research~\cite{blackbox, liu2024your}, as ``seeds'' for the model to generate code snippets for our experiments. Specifically, this dataset contains 164 programming problems with unit tests. Each problem has a problem description, a function signature, and several manually created test cases. On average, there are about eight tests per problem to evaluate the correctness of the generated code. 
For thorough evaluation, 
\tool queries the models to generate code in two popular programming languages, Java and Python, for each problem. Then, \tool examines how the model understands code semantics in both languages.

\tool considers three SOTA open-source code LLMs, including \deepseek~\cite{deepseek-coder}, \codellama~\cite{codellma}, and \magiccoder~\cite{magicoder}. In addition, we also evaluate the capability of a commercial model, \chatgpt~\cite{gpt}, which was the first model deployed for ChatGPT. For \deepseek\footnote{https://huggingface.co/deepseek-ai/deepseek-coder-6.7b-instruct}, \codellama\footnote{https://huggingface.co/codellama/CodeLlama-7b-Instruct-hf}, and \magiccoder\footnote{https://huggingface.co/ise-uiuc/Magicoder-DS-6.7B}, \tool adapts their instructed models officially deployed on \textit{HuggingFace} without further training. For \chatgpt\footnote{https://platform.openai.com/docs/models/gpt-3-5-turbo}, \tool utilizes their provided APIs. Note that to reduce the threat of randomness during the generation process of the LLMs, we set the temperature to $0$ for all the studied models. The performance of these models (Pass@1) on the \textit{HumanEval} benchmark is shown in Table~\ref{tab:pass@1}. These generated code snippets are utilized as inputs to evaluate the models in the code understanding tasks.

\begin{table}[!htp]\centering
\caption{Code LLMs' performance (Pass@1) on \textit{HumanEval}}
\label{tab:pass@1}
\begin{tabular}{l|c|c|cc}\toprule
            &Size    &Java   &Python \\\midrule
\codellama  & 7B      &29.7\% &36.0\% \\
\chatgpt    &  175B     &58.2\% &65.9\% \\
\deepseek   & 6.7B      &61.4\% &70.7\% \\
\magiccoder & 6.7B      &60.1\% &73.2\% \\
\bottomrule
\end{tabular}
\end{table}

\textbf{RQ1. \textit{Robustness and Sensitivity Analysis}.} We study how code LLMs are robust to the \textit{SP} transformations and how they are sensitive to the \textit{SNP} transformations. In this experiment, we apply each transformation operator on each code snippet and then observe the variations of the models' responses on three code understanding tasks: \textit{code summarization, method name prediction}, and \textit{output prediction}. For each task, the \textbf{$Robustness$} of a code LLM is measured by the \textbf{average} robustness of the model to all the \textit{SP} transformation operators. The \textbf{$Sensitivity$} of the model is calculated by the  \textbf{average} sensitivity of the model to all the \textit{SNP} transformation operators.

\textbf{RQ2. \textit{Model Size Analysis}.} We investigate the impact of the number of parameters on the model's capability of understanding code semantics. \tool conducts experiments on different variants of \deepseek  (1.3B, 6.7B, and 33B parameters) and studies how the number of parameters affects models' robustness and sensitivity in different code understanding tasks.

\textbf{RQ3. \textit{Correctness Correlation Analysis}.} We study the correlation between the code semantic understandability of the models and their code generation performance. We argue that the models better understand the correct code than the incorrect one. Thus, for the correct code, the models perform more robustly to the \textit{SP} transformations and more sensitively to the \textit{SNP} transformation than for the incorrect one.

\textbf{RQ4. \textit{Semantic Inference Analysis}.} We aim to evaluate whether the code LLMs can understand and correctly infer the code semantics regarding control and data dependencies as the program analyzer does. We prompt the models to list all the pairs of code statements which have control/data dependence for each code snippet. Then, we compare the performance of the models with the control/data dependencies identified by a program analyzer. In this work, we employ Joern~\cite{cpg}, which is widely used in SE-related studies, as a reference program analyzer.

\section{Experimental Results}
\label{sec:results}

\subsection{RQ1. Robustness and Sensitivity Analysis}

\subsubsection{Code Summarization}

\begin{table*}[!htp]\centering
\caption{RQ1. The $Robustness$ and $Sensitivity$ of code LLMs to the code transformations in the code summarization task}
\label{tab:code_summarization}
\resizebox{\textwidth}{!}{%
\begin{tabular}{l|l|rrrr|rrrrr}\toprule
\multicolumn{2}{c}{\multirow{2}{*}{}} &\multicolumn{4}{|c|}{\textbf{Java}} &\multicolumn{4}{c}{\textbf{Python}} \\\midrule
& &\codellama &\chatgpt &\deepseek &\magiccoder &\codellama  &\chatgpt &\deepseek &\magiccoder \\ \midrule
\multirow{2}{*}{\textbf{Semantic Sim.}} &\textbf{\robust} &0.76 &0.95 &0.93 &0.92 &0.91 &0.90 &0.92 &0.92 \\
&\textbf{\sensitive} &0.17 &0.07 &0.09 &0.09 &0.08 &0.10 &0.09 &0.09 \\ \midrule
\multirow{2}{*}{\textbf{Lexical Sim.}} &\textbf{\robust } &0.77 &0.82 &0.85 &0.85 &0.85 &0.75 &0.88 &0.84 \\
&\textbf{\sensitive} &0.19 &0.22 &0.16 &0.18 &0.16 &0.26 &0.16 &0.17\\
\bottomrule
\end{tabular}
}
\end{table*}


Table~\ref{tab:code_summarization} presents the average $Robustness$ and $Sensitivity$ of code LLMs to the code transformations in the code summarization task. \textit{Code LLMs perform robustly to both SP and SNP transformations}. 
Particularly, regarding semantic similarity, the average $Robustness$ of the models to the \textit{SP} transformations is about 0.9. The average $Sensitivity$ of the models to the \textit{SNP} transformations is about 0.1. In other words, the similarities between the summaries of the original code and those of the semantic equivalent transformed code are 0.9. Meanwhile, the differences between the summaries of the original code and those of the semantic non-equivalent transformed code are only 0.1. This indicates that the code LLMs produce similar summaries for the original code and the transformed code \textit{\textbf{regardless}} of the impact of the transformations on the code semantics.

\begin{figure}[]
\centering
\begin{subfigure}{0.5\textwidth}
\centering
 \includegraphics[width=0.85\columnwidth]{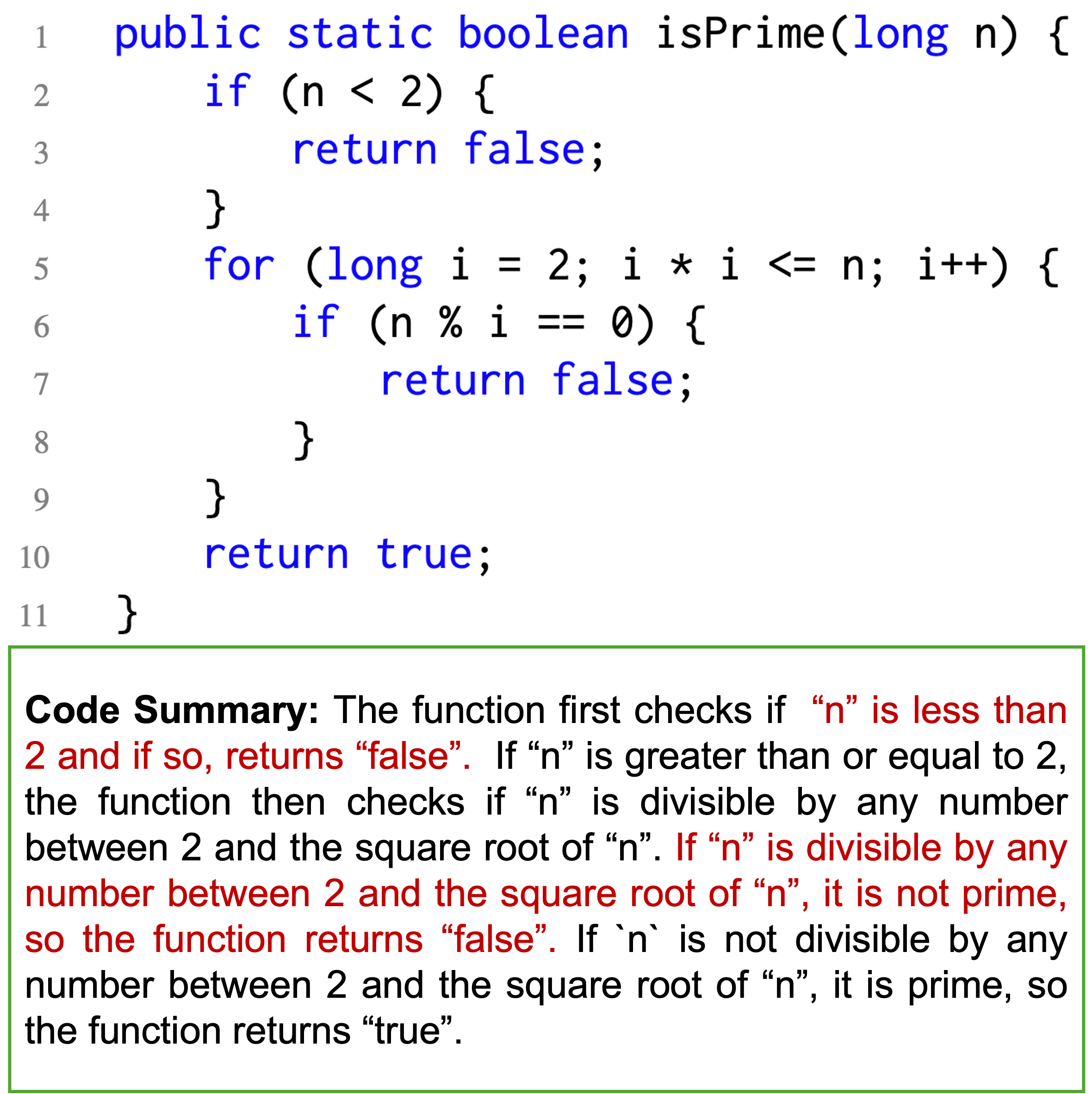}
\caption{The summary of the original code}
\label{fig:summary_ex1}
\end{subfigure}\\
\begin{subfigure}{0.5\textwidth}
\centering
 \includegraphics[width=0.85\columnwidth]{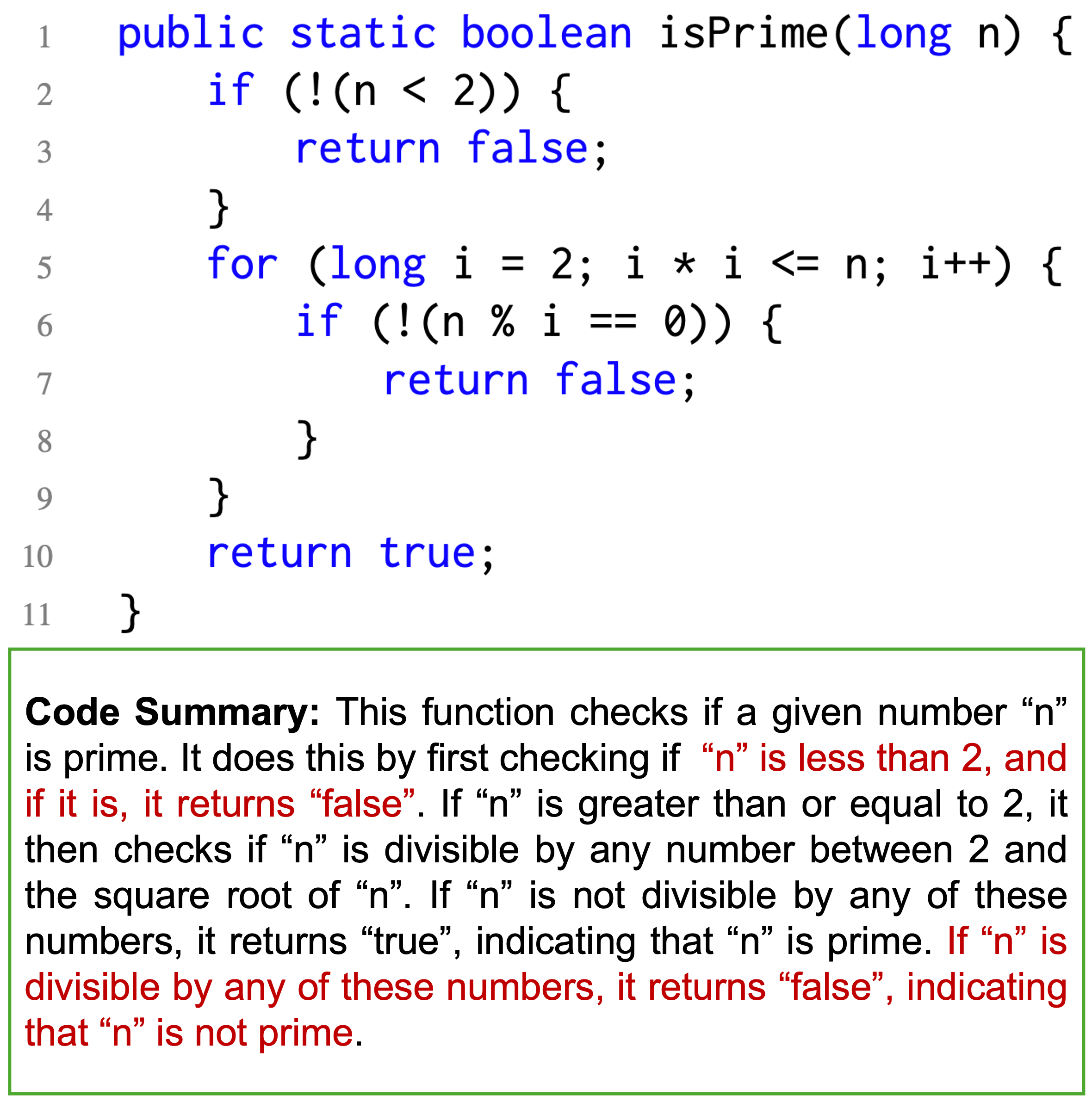}
\caption{The summary of the code transformed by negating the condition}
\label{fig:summary_ex2}
\end{subfigure}

\caption{The summaries produced by \codellama for the original code and the  transformed one (HumanEval\_31\_is\_prime)}
\label{fig:summary_example}
\end{figure}

Figure~\ref{fig:summary_example} shows an example about the summaries produced by \codellama for its generated code and for the corresponding transformed code. 
The original generated code is to check if an input $n$ is a prime number. As shown, the summary of the original code (Figure~\ref{fig:summary_ex1}) and that of the transformed code, which is created by negating the conditions at lines $2$ and $6$ (Figure~\ref{fig:summary_ex2}), are pretty similar.
%
The transformations completely change the semantics of the original code. Specifically, the transformed code returns \texttt{false} for all the numbers greater than or equal to $2$. This means that the transformed code (Figure~\ref{fig:summary_ex2}) no longer checks for prime numbers. However, the general structures of the original code and the transformed one are quite similar. As a result, \codellama produces similar summaries for these two methods regardless their functionality differences.

In general, code LLMs are not explicitly trained to capture the program dependencies, the underlying semantics, or the operational logic of code. Instead, they are trained on a large corpus of code in form of text data to implicitly learn the patterns and relationships in the data. When generating code summaries, the models could rely on the overall structures and the surrounding contexts rather than deeply understanding the underlying semantics of each code statement in the given code snippets. 
Although the original code and the transformed code could be different or similar regarding their semantics, their general structures are still similar. Therefore, the models could produce similar summaries for them. For instance, the transformations in Figure~\ref{fig:summary_ex2} make the behaviors of the program totally different from those of the original one (Figure~\ref{fig:summary_ex1}). However, the general patterns of these two code snippets are similar, which led \codellama to result in similar summaries.
That is a reason why code LLMs demonstrate robust performance against all transformations in the task of code summarization.

\begin{gtheorem}
\textbf{Finding 1}: For the task of code summarization, code LLMs are robust to both SP and SNP transformations.
\end{gtheorem}

\subsubsection{Method Name Prediction}

\begin{table*}[!htp]\centering
\caption{RQ1. The $Robustness$ and $Sensitivity$ of Code LLMs to the code transformations in the method name prediction task}
\label{tab:method_name}
\resizebox{\textwidth}{!}{%
\begin{tabular}{l|l|rrrr|rrrrr}\toprule
\multicolumn{2}{c}{\multirow{2}{*}{}} &\multicolumn{4}{|c|}{\textbf{Java}} &\multicolumn{4}{c}{\textbf{Python}} \\\midrule
& &\codellama &\chatgpt &\deepseek &\magiccoder &\codellama &\chatgpt &\deepseek &\magiccoder \\\midrule
\multirow{2}{*}{\textbf{Exactly Match}} &\textbf{\robust } &0.59 &0.53 &0.65 &0.71 &0.44 &0.40 &0.53 &0.56 \\
&\textbf{\sensitive} &0.50 &0.62 &0.54 &0.52 &0.58 &0.58 &0.65 &0.56 \\ \midrule
\multirow{2}{*}{\textbf{F1-Score}} &\textbf{\robust } &0.75 &0.79 &0.82 &0.85 &0.65 &0.69 &0.76 &0.76\\
&\textbf{\sensitive} &0.30 &0.32 &0.33 &0.31 &0.39 &0.30 &0.36 &0.32 \\
\bottomrule
\end{tabular}
}
\end{table*}

\textit{Compared to the task of code summarization, code LLMs are less robust and more sensitive to both kinds of transformations in the task of method name prediction}, see Table~\ref{tab:method_name}. For instance, the predicted method names of 62\% \textit{SP} transformed Java code and 48\%  \textit{SP} transformed Python code are identical to those of the original code snippets. Meanwhile, by \textit{SNP} transformations, 54\% of transformed Java code and 59\% of transformed Python code have the predicted names different from that of the original code.
This could indicate that 
in the method name prediction task,
code LLMs are more perceptive to modifications and respond differently to original and transformed code.

\begin{figure}[]
\centering
\begin{subfigure}{0.5\textwidth}
\centering
\includegraphics[width=0.85\columnwidth]{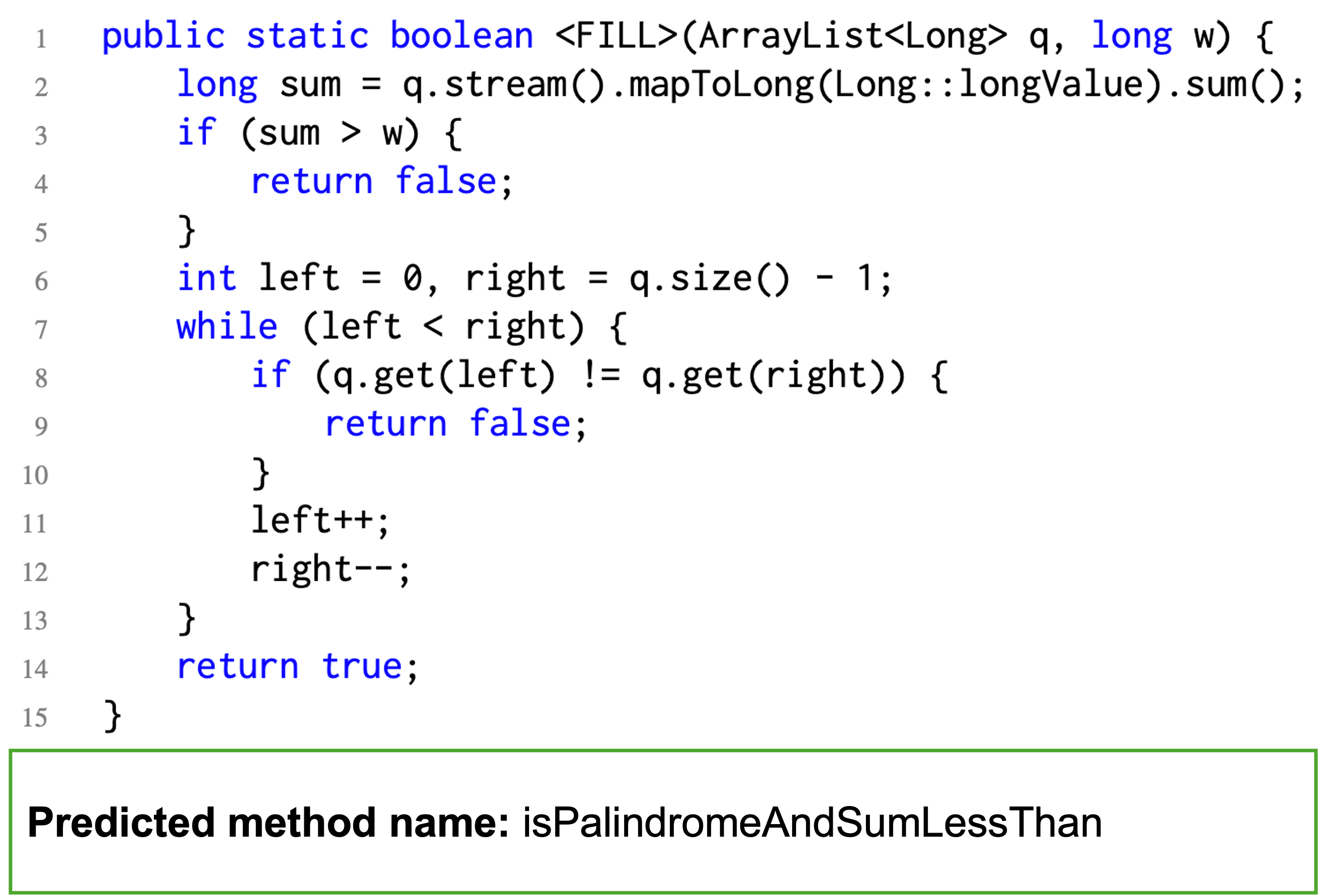}
\caption{The predicted method name of the original code}
\label{fig:name_ex1}
\end{subfigure}\\
\begin{subfigure}{0.5\textwidth}
\centering
\includegraphics[width=0.85\columnwidth]{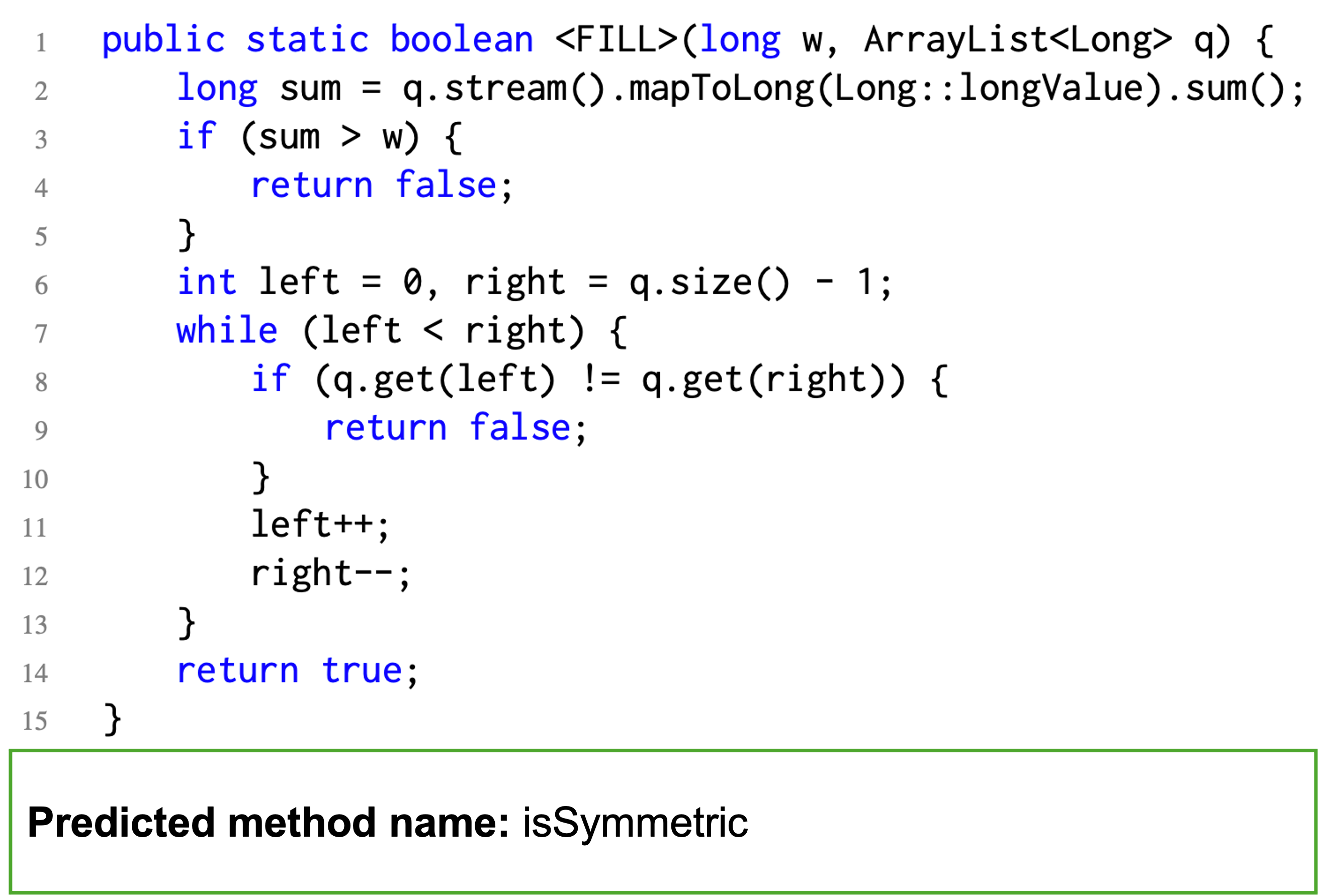}
\caption{The predicted method name of the code transformed by reordering parameters}
\label{fig:name_ex2}
\end{subfigure}\\

\caption{The method name predicted by \magiccoder for the original generated code and the transformed one (HumanEval\_72\_will\_it\_fly)}
\label{fig:name_example}
\end{figure}

Figure~\ref{fig:name_example} shows the method name predicted by \magiccoder for its original generated code and the corresponding transformed code. As seen in Figure~\ref{fig:name_ex1}, the code originally generated by \magiccoder for problem of HummanEval\_72\_will\_it\_fly is correct, which correctly returns \texttt{true} if an array list \texttt{q} will fly. Specifically, \texttt{q} will fly if it is a palindromic array list and the sum of its elements is less than or equal to a weight \texttt{w}.

For the original code in Figure~\ref{fig:name_ex1}, \magiccoder suggested the method name \texttt{isPalindromeAndSumLessThen}, which reasonably describes the functionality of the given code. However, after \textit{reordering the parameters} (Figure~\ref{fig:name_ex2}), \magiccoder suggests a \textit{completely} different name, \texttt{isSymmetric} even though this transformation does not impact the functionality of the code. This shows that \magiccoder is not robust enough to preserve the recommended name for the semantic equivalent transformed code. 

For the similarity at the sub-token level (F1-Score), code LLMs show their high $Robustness$ to the \textit{SP} transformations but low $Sensitivity$ to the \textit{SNP} transformations. In particular, the average $Robustness$  in terms of F1-Score for Java code is 0.80, and the corresponding figure for Python code is 0.72. Meanwhile, the $Sensitivity$ is about 0.31 and 0.34 for Java and Python, respectively. This indicates that code LLMs can generate similar method names for the original code and \textit{SP} transformed code. However, for \textit{SNP} transformed code, the predicted names are not considerably distinguishable from that of the original code. 
These results suggest that at some degree, code LLMs can capture the code semantics and realize the modifications affecting code meaning. 
However, the effects on the generated outputs (methods' names) are not significant.

\begin{gtheorem}
\textbf{Finding 2}: Compared to the task of code summarization, code LLMs are more sensitive to all the transformations in the task of method name prediction. For this task, code LLMs can somewhat capture the code semantics and realize the SNP transformations to vary their responses. However, the effects on the outputs are not significant. 
\end{gtheorem}

\subsubsection{Output Prediction}

\begin{table*}[!htp]\centering
\caption{RQ1. The $Robustness$ and $Sensitivity$ of code LLMs to the code transformations in the output prediction task}
\label{tab:output}
\resizebox{\textwidth}{!}{%
\begin{tabular}{l|rrrr|rrrrr}\toprule
\multirow{2}{*}{} &\multicolumn{4}{|c|}{\textbf{Java}} &\multicolumn{4}{c}{\textbf{Python}} \\\midrule
&\codellama &\chatgpt &\deepseek &\magiccoder &\codellama &\chatgpt &\deepseek &\magiccoder \\ \midrule
\textbf{\robust } &0.66 &0.65 &0.69 &0.69 &0.67 &0.62 &0.53 &0.60 \\
\textbf{\sensitive} &0.40 &0.47 &0.48 &0.47 &0.37 &0.47 &0.57 &0.55 \\
\bottomrule
\end{tabular}
}
\end{table*}
Table~\ref{tab:output} shows the average $Robustness$ and $Sensitivity$ of code LLMs for the task of \textit{output prediction}.
The results illustrate the significant differences between the outputs predicted for the original code and the transformed code. 
Specifically, only 64\% of the cases are predicted to have equal outputs in the original code and the \textit{SP} transformed code. Meanwhile, 47\% of the cases are predicted to have different outputs in the original code and the \textit{SNP} transformed code.
Overall, \textit{code LLMs are quite sensitive to all the transformations in predicting outputs.}



As seen the results from the three studied tasks, \textit{code LLMs tend to be more robust to the \textit{SP} transformations than to be sensitive to the \textit{SNP} transformations}.
Indeed, for the tasks of method name prediction and output prediction, code LLMs better recognize of the semantic equivalence caused by the \textit{SP} transformations rather than the semantic inequivalence by the \textit{SNP} transformations. For example, in Table~\ref{tab:output}, for transformed Java code, LLMs could preserve equal outputs for about 67\% of cases in the \textit{SP} transformed code. However, for \textit{SNP} ones, it can recognize the differences and produce different outputs for only 45\% of cases. 


\begin{gtheorem}
\textbf{Finding 3}: Code LLMs are better at the property of $Robustness$ to the \textit{SP} transformations than the property of $Sensitivity$ to the \textit{SNP} transformations.
\end{gtheorem}


\textbf{Transformation Operator Analysis:}
Table~\ref{tab:transformation_operators} presents the details $Robustness$ and $Sensitivity$ of \deepseek to different transformation operators on each code understanding task\footnote{The results of the other LLMs could be found on our website.}. 
%
%
In our experiment, we found that for all the tasks, \textit{Renaming Variable} creates the strongest impacts on the models among the \textit{SP} operators although they have no impact on the functionalities of the given code snippets. 
For instance, the summaries of the original code and its renamed variable versions are 88\% semantically similar which is much lower than the average similarity values of the other \textit{SP} transformation operators, 94\%.
For the task of \textit{method name prediction}, after renaming variables, only 33\% of the code snippets still have the same predicted names as the original ones, which is more than \textit{two times} lower than the average results of the other \textit{SP} operators.

In practice, the variable names often convey rich information for code understanding. Additionally, language models often rely heavily on context to generate predictions or summaries. Thus, changing variable names can disrupt this context and lead code LLMs to alter their understanding of the code's intent and produce different outputs.

\begin{table*}[!htp]\centering
\caption{RQ1. The $Robustness$ and $Sensitivity$ of code LLMs to different transformation operators}
\label{tab:transformation_operators}
\begin{tabular}{l|l|l|rr|rr|rrr}\toprule
\multicolumn{3}{c}{\multirow{2}{*}{\textbf{}}} &\multicolumn{2}{|c|}{\textbf{\begin{tabular}[c]{@{}c@{}}Code\\ Summarization \end{tabular}}} &\multicolumn{2}{c|}{\textbf{\begin{tabular}[c]{@{}c@{}}Method Name \\ Prediction \end{tabular}}} &\multicolumn{2}{c}{\textbf{\begin{tabular}[c]{@{}c@{}}Output\\ Prediction \end{tabular}}} \\\midrule
& & &Java &Python &Java &Python &Java &Python \\ \midrule
\multirow{4}{*}{\textbf{\robust }} &\multirow{2}{*}{\begin{tabular}[c]{@{}l@{}}Control\\ Dependence \end{tabular}} &Convert For/While &0.95 &0.92 &0.72 &0.60 &0.78 &0.57 \\\cmidrule{3-9}
& &Flip IfElse &0.94 &0.95 &0.80 &0.53 &0.68 &0.52 \\\cmidrule{2-9}
&\multirow{2}{*}{\begin{tabular}[c]{@{}l@{}}Data\\ Dependence \end{tabular} } &Rename Variable &0.88 &0.85 &0.34 &0.32 &0.62 &0.51 \\\cmidrule{3-9}
& &Reorder Parameters &0.93 &0.95 &0.72 &0.66 &- &- \\ \midrule
\multirow{4}{*}{\textbf{\sensitive }} &\multirow{2}{*}{\begin{tabular}[c]{@{}l@{}}Control\\ Dependence \end{tabular}} &\begin{tabular}[c]{@{}l@{}}Negate Relational\\ Condition \end{tabular} &0.06 &0.05 &0.55 &0.62 &0.38 &0.50 \\\cmidrule{3-9}
& &\begin{tabular}[c]{@{}l@{}}Remove Conditional\\ Statement \end{tabular} &0.12 &0.11 &0.61 &0.72 &0.52 &0.62 \\\cmidrule{2-9}
&\multirow{2}{*}{\begin{tabular}[c]{@{}l@{}}Data\\ Dependence \end{tabular}} &\begin{tabular}[c]{@{}l@{}}Replace Arithmetic\\ Operator \end{tabular} &0.08 &0.09 &0.51 &0.65 &0.51 &0.62 \\\cmidrule{3-9}
& &\begin{tabular}[c]{@{}l@{}}Remove Def\\ Statement \end{tabular} &0.09 &0.09 &0.49 &0.61 &0.50 &0.54 \\
\bottomrule
\end{tabular}
\end{table*}

Meanwhile, among the \textit{SNP} transformation operators, \textit{removing conditional statements} is the most sensitive for the code LLMs. 
This operator leads to the most significant differences in the responses of code LLMs compared to their responses to the original code. For example, \deepseek changed its predicted method names in 62\% python programs when the relation conditions were negated. While by removing conditional statements, this figure increased to 72\%. Indeed, this transformation operator alters the code's structure and texture more extensively than the other operators. This makes the differences between the transformed and the original code more apparent to the models, which affects their predictions.

\begin{gtheorem}
\textbf{Finding 4}: The $Robustness$ and $Sensitivity$ of code LLMs to the code transformations vary for each code understanding task and each transformation operator. For all the transformations in general, code LLMs are the most robust in the code summarization task and the most sensitive in the output prediction task. For each task, the rename variable operator has the most considerable impact on the models' outputs.
\end{gtheorem}

\subsection{RQ2. Model Size Analysis}

\begin{figure}[]
\centering
\begin{subfigure}{.5\textwidth}
\centering
\includegraphics[width=1\linewidth]{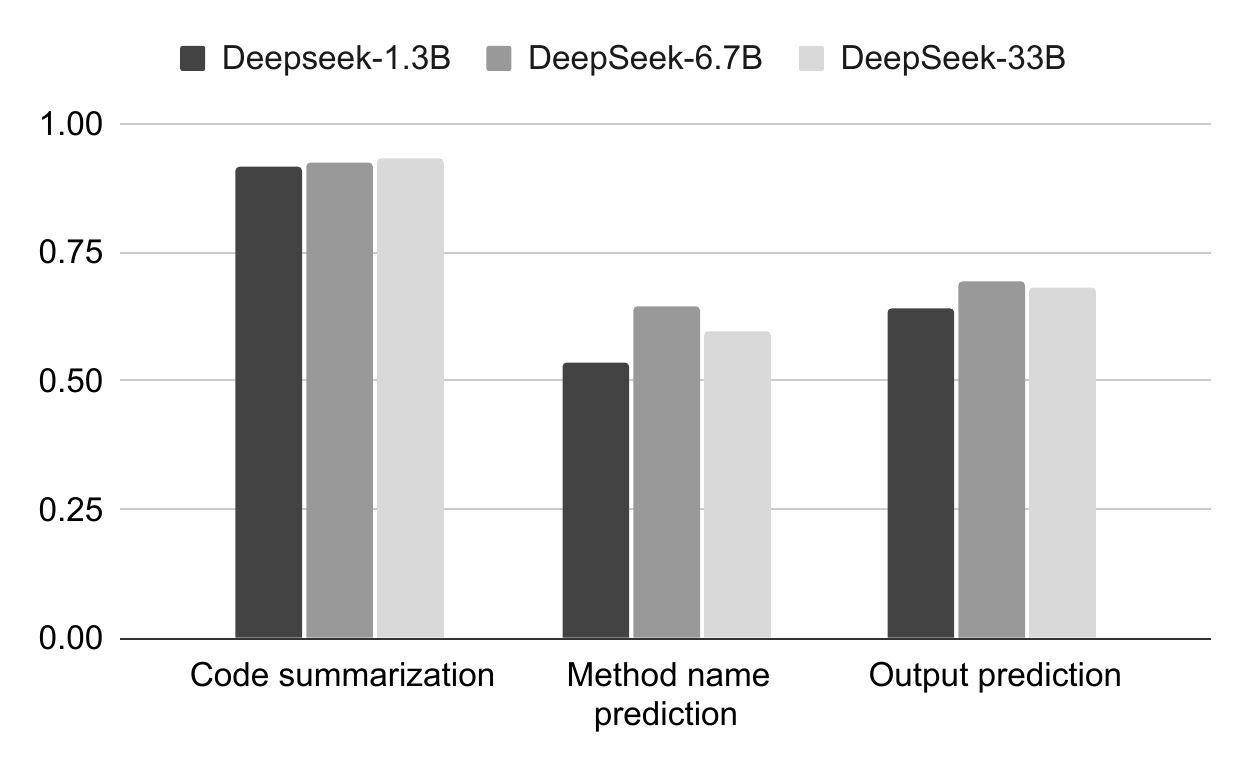}
\caption{\robust}
\label{fig:modelsize_robustness}
\end{subfigure}%

\begin{subfigure}{.5\textwidth}
\centering
\includegraphics[width=1\linewidth]{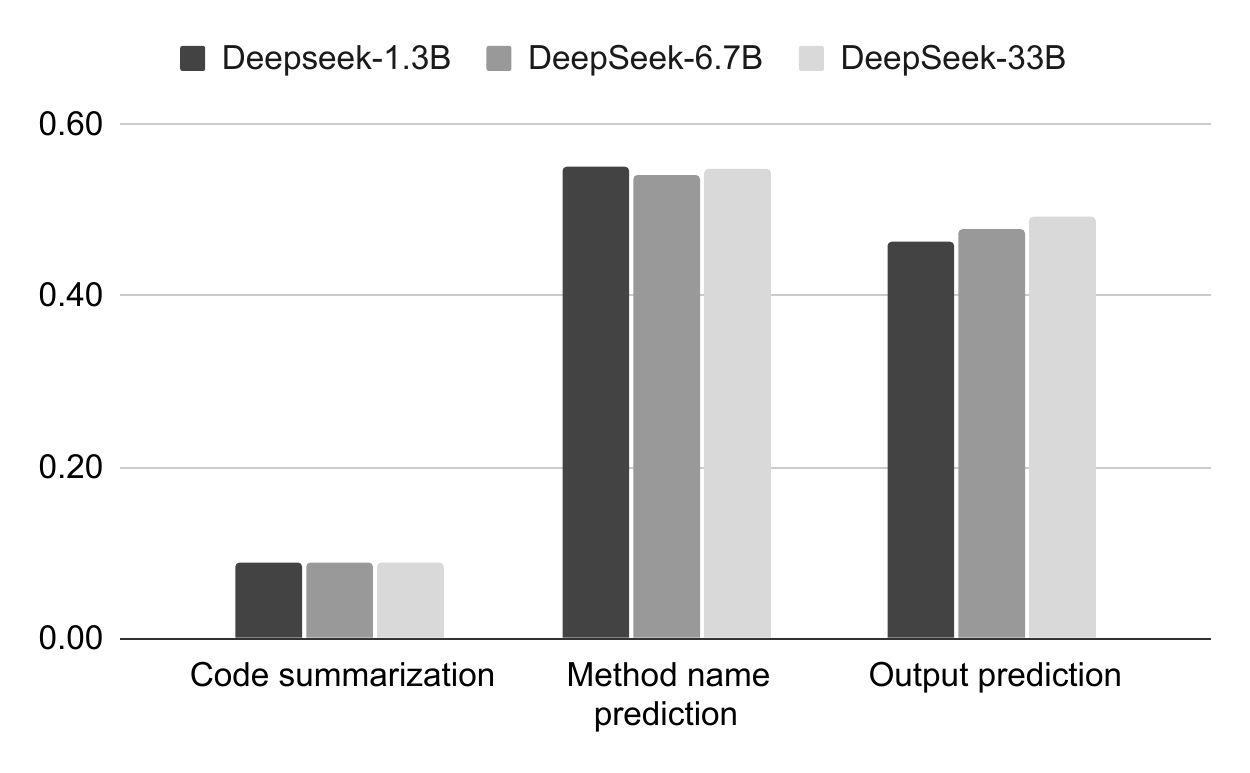}
\caption{\sensitive}
\label{fig:modelsize_sensitivity}
\end{subfigure}
\caption{RQ3. The impacts of model sizes on \deepseek's capability in capturing code semantics}
\label{fig:modelsize}
\end{figure}

Figure~\ref{fig:modelsize} shows the impact of the model sizes on the $Robustness$ and $Sensitivity$ of \deepseek to the transformations.
Overall, \textit{the Robustness and Sensitivity of code LLMs to the transformations are greatly affected by the code understanding tasks rather than the number of the models' parameters}.

In particular, for all the studied variants of \deepseek, the $Robustness$ and $Sensitivity$ of the models in the \textit{code summarization} task are around 0.92 and 0.09, respectively. This means that regardless of the number of parameters, \deepseek produces 92\% similar summaries for the original code and the \textit{SP} transformed code, while the differences between the summaries for the original code and the \textit{SNP} transformed code are only 9\%. These figures are quite stable for the variants with different parameter sizes of \deepseek.

Meanwhile, for \textit{output prediction}, the $Robustness$ values range from 0.64 to 0.69. These figures are not considerably different among model variants, yet much lower than that of the task of code summarization. Similarly, all the studied variants of \deepseek have the similar $Sensitivity$ values for the task of \textit{output prediction}, and these figures are significantly higher than those of code summarization.


\begin{gtheorem}
\textbf{Finding 5}: The $Robustness$ and $Sensitivity$ of code LLMs to the transformations are not significantly affected by the number of the models' parameters.
\end{gtheorem}
\subsection{RQ3. Correctness Correlation}

\begin{figure}[]
\centering
\begin{subfigure}{.5\textwidth}
\centering
\includegraphics[width=1\linewidth]{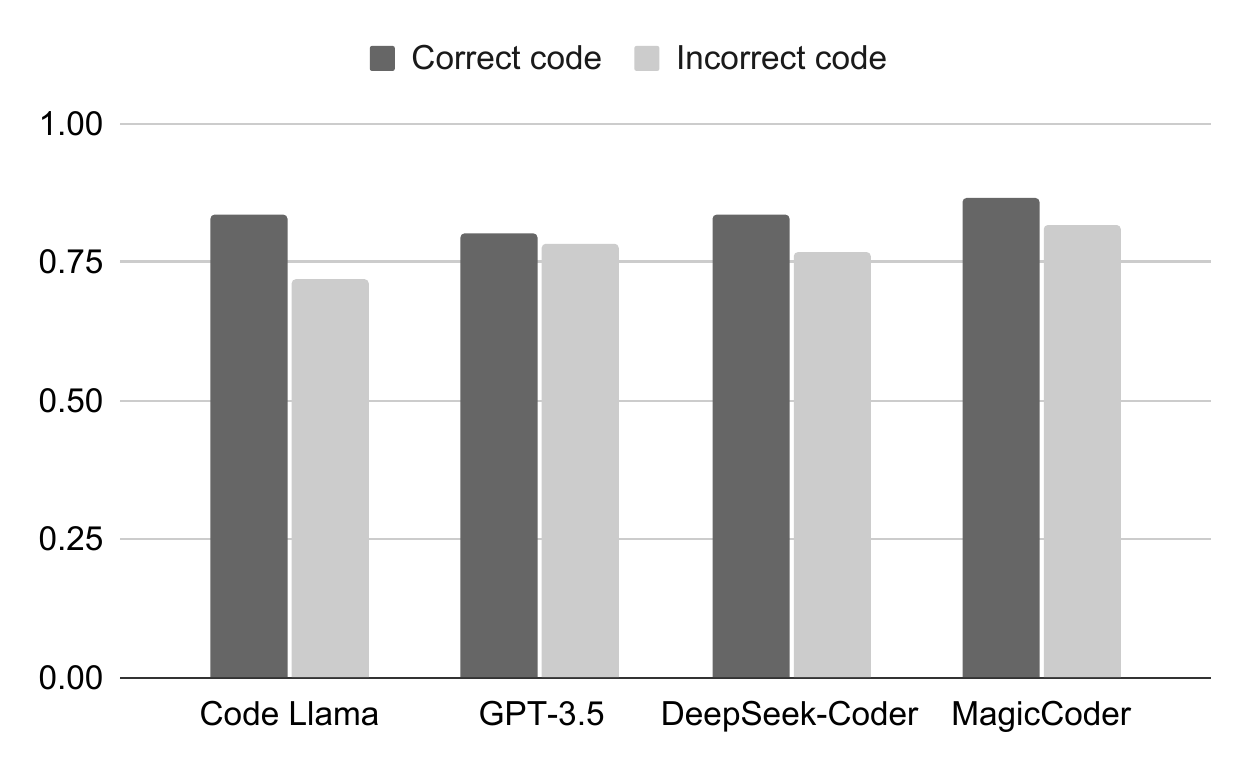}
\caption{\robust}
\label{fig:correctness_robustness}
\end{subfigure}%

\begin{subfigure}{.5\textwidth}
\centering
\includegraphics[width=1\linewidth]{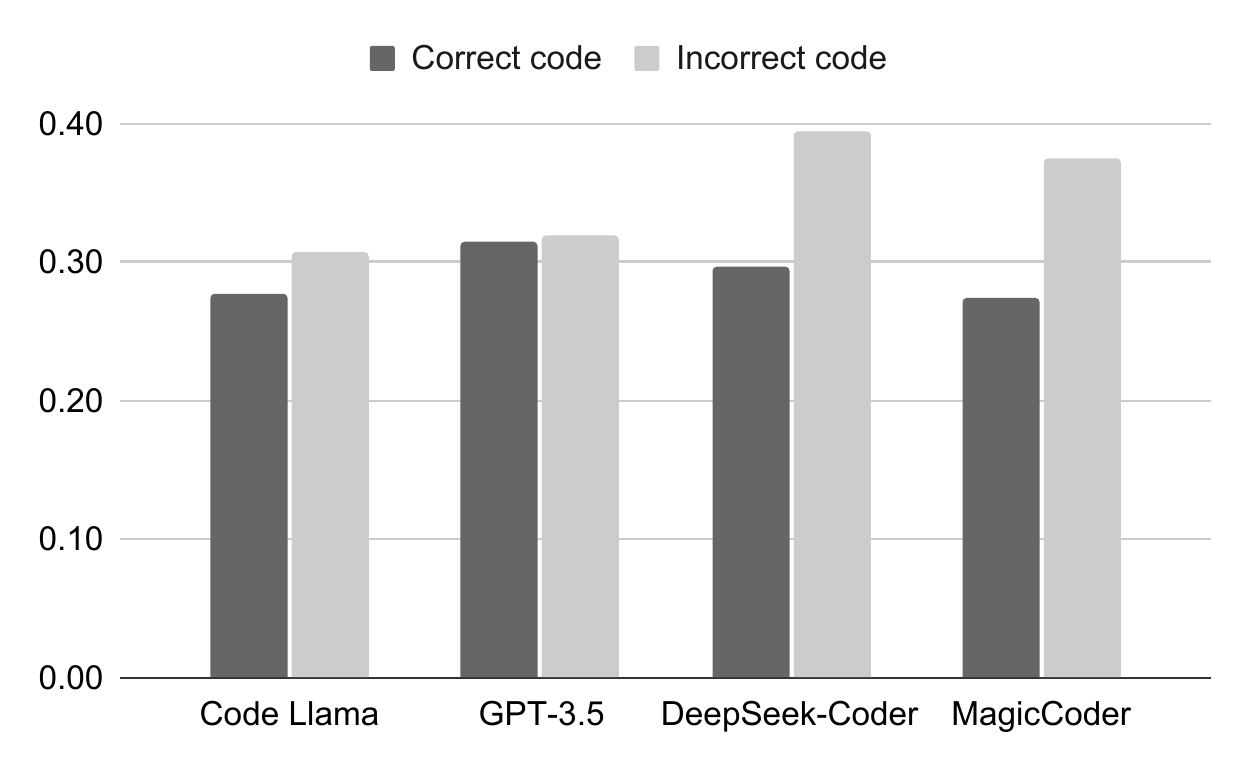}
\caption{\sensitive }
\label{fig:correctness_sensitivity}
\end{subfigure}
\caption{RQ4. The semantic understandability of the code LLMs and the correctness of their generated code}
\label{fig:correctness}
\end{figure}

Figure~\ref{fig:correctness} shows the $Robustness$ and $Sensitivity$ of Code LLMs to the transformations of the correct and incorrect generated code in the task of method name prediction.
For the correct code, 84\% of the semantic equivalent code snippets are suggested similar names by code LLMs. Meanwhile, the corresponding figure for the incorrect code is only 77\%. For the semantic non-equivalent code snippets, the cases with different predicted names account for 30\% for the correct programs and 35\% for the incorrect programs.
This demonstrates that \textit{code LLMs are more robust to the transformations of the corrected code and more sensitive to the transformations of incorrect code.} 

One of the reasons could be that the incorrect code may deviate significantly from the patterns present in the training data. When applying the transformations to incorrect code, it might introduce changes that are further different from the patterns that code LLMs have learned. Consequently, capturing the meaning of these transformed code snippets becomes more challenging for the models. Therefore, their performance is less stable for the incorrect code and its transformed versions.


\begin{gtheorem}
\textbf{Finding 7}: The models are more robust to the transformations of the correct code and more sensitive to the transformations of incorrect code.
\end{gtheorem}
\subsection{RQ4. Semantic Inference Analysis}

\begin{table*}[!htp]\centering
\caption{RQ2. Performance of code LLMs in inferring control/data dependence of the code snippets}
\label{tab:dependence_concepts}
\begin{tabular}{l|l|rrr|rrrr}\toprule
\multirow{2}{*}{Language} &\multirow{2}{*}{Model} &\multicolumn{3}{c|}{\textbf{Control dependence}} &\multicolumn{3}{c}{\textbf{Data dependence}} \\\cmidrule{3-8}
& &Precision &Recall &F1-Score &Precision &Recall &F1-Score \\ \midrule
\multirow{4}{*}{Java} &\codellama &0.25 &0.24 &0.24 &0.02 &0.01 &0.02 \\
&\chatgpt &0.29 &0.29 &0.29 &0.11 &0.05 &0.06 \\
&\deepseek &0.31 &0.31 &0.29 &0.12 &0.04 &0.05 \\
&\magiccoder &0.30 &0.29 &0.28 &0.03 &0.01 &0.01 \\ \midrule
\multirow{4}{*}{Python} &\codellama &0.44 &0.42 &0.43 &0.01 &0.01 &0.01 \\ 
&\chatgpt &0.56 &0.54 &0.55 &0.03 &0.01 &0.02 \\
&\deepseek &0.39 &0.33 &0.36 &0.15 &0.05 &0.07 \\
&\magiccoder &0.32 &0.31 &0.31 &0.05 &0.02 &0.03 \\
\bottomrule
\end{tabular}
\end{table*}

Table~\ref{tab:dependence_concepts} shows the prediction performance of Code LLMs in predicting pairs of statements (\textit{pairs} for short) that have control/data dependence relationships in a given code. 
%
%
Specifically, \textit{Precision} specifies the ratio of pairs that really have a control/data dependence relationship among the pairs predicted by the models. 
Meanwhile, \textit{Recall} is the percentage of pairs having a control/data dependence relationship correctly predicted by the models. 
\textit{F1-Score} is the harmonic mean of precision and recall.

As seen in Table~\ref{tab:dependence_concepts}, \textit{all the studied LLMs can capture the control dependence much better than the data dependence}. For example, \deepseek precisely predicted 30-40\% pairs of control-dependent statements. Meanwhile, the corresponding figures for data dependence are only 12-15\%. Especially, \chatgpt can recognize up to 54\% of pairs of control-dependent Python statements, but it totally failed to recognize the data dependence, only 1\% in $Recall$. 

One of the reasons why code LLMs are more effective at capturing control dependence than data dependence is that the control dependence is often expressed by clear markers in code. Specifically, control dependence refers to the relationship between code statements to determine which statements are executed under certain conditions. This relationship is explicitly expressed in structures like \texttt{if-else} blocks or loops. This makes it easier for code LLMs to identify and understand this kind of relationship. Meanwhile, data dependence is illustrated via variable assignments, usages, or transformations. It can be subtle for code LLMs and requires a deep understanding of the code operation to capture data dependence accurately.

Furthermore, \codellama and \chatgpt perform better on Python than on Java. Meanwhile, \deepseek and \magiccoder are stable on both Python and Java. Specifically, for both Java and Python, \deepseek and \magiccoder achieved a precision of about 30\% in control dependence prediction. The performance of \codellama and \chatgpt on control dependence prediction on Python is 44\% and 56\%, nearly double their performance on Java code.


\begin{figure}
    \centering
    \includegraphics[width=0.85\columnwidth]{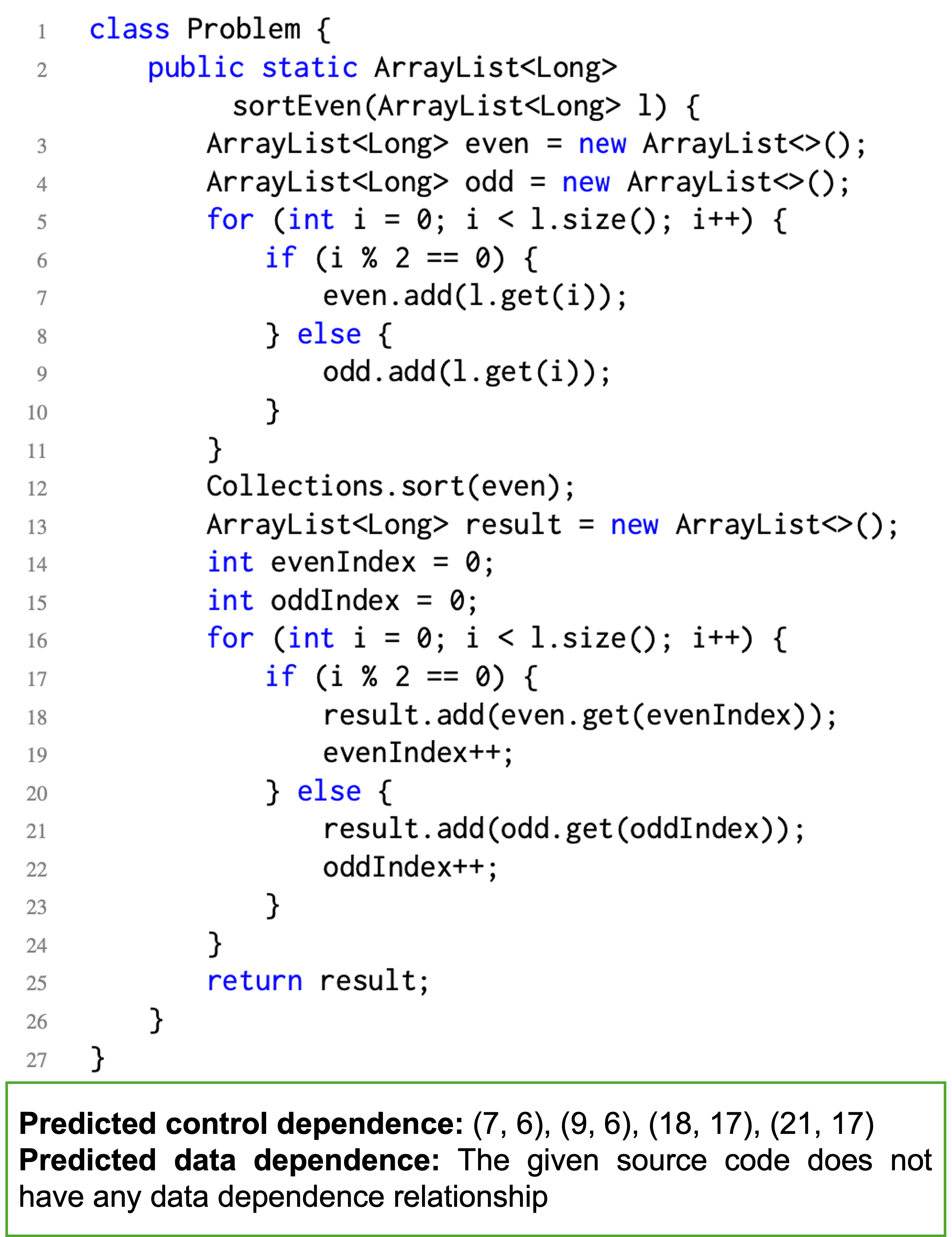}
    \caption{Example of the control and dependence predicted by \magiccoder, Problem ID: HumanEval\_37\_sort\_even}
    \label{fig:control_dependence}
\end{figure}

Figure~\ref{fig:control_dependence} shows an example of the control- and data-dependencies predicted by \magiccoder. As seen, \magiccoder correctly captured the statements that are dependent on the conditional statements (line 6 and line 17). In particular, it correctly predicted the statements at line 7 and line 9 are control-dependent on line 6. Also, the statements at line 18 and line 21 are control dependent on the statement at line 17. Although it still missed some other statements (line 19 or line 22), which are also control-dependent on the statement at line 17 or the control dependence caused by the for-loop (line 5 and line 16), all of the predicted pairs of control-dependent statements are correct. However, for data dependence of this code snippet, \magiccoder failed to capture and cannot predict any pairs of data-dependent statements. Indeed, the data dependence pattern in this source code is expressed via argument passing instead of simply assigning variables. This could be quite complex for the model to capture precisely.


\begin{gtheorem}
\textbf{Finding 8}: Code LLMs have limited ability to directly infer the control dependencies in source code. Especially, the models still 
severely suffer from capturing data dependencies.
\end{gtheorem}




\subsection{Threats to Validity}

As for any empirical study, there are several threats to the validity of our results and conclusions. First, a potential threat lies in the design of the prompts. 
To mitigate this threat, we carefully follow the prompt templates of the studied models to design and optimize the prompts for each code understanding task. Second, there is a certain level of randomness of the studied code LLMs in the generation procedure, which could affect the models' performance. 
To reduce this threat, we set the temperate value to 0 to minimize the randomness. 
Another threat could lie in the selection code LLMs. To mitigate this threat, we chose diverse and representative models, including both open-source and commercial ones. We also conducted experiments on different model sizes.  
In addition, one threat could come from the evaluation procedure, i.e., different performance metrics may exhibit different biases. To mitigate this threat, for each code understanding task, we measure the similarity of the models' outputs for the original code and the transformed code at different levels. For example, for code summarization, the similarity is measured at both semantic and lexical levels.
\section{Conclusion}
\label{sec:conclusion}


In this paper, we introduce \tool, a novel framework designed to systematically and empirically evaluate the code semantic understanding of code LLMs. Specifically, \tool systematically introduces controlled modifications into the input code and examines the models' responses. For a given code snippet $c$,  if a code LLM understands the semantics of $c$, the model should produce consistent and robust outputs for $c$ and its semantic equivalent variants while generating different responses for $c$ and its semantic non-equivalent variants. Our experimental results on three representative code understanding tasks, code summarization, method name prediction, and output prediction, reveal that the robustness and sensitivity of the state-of-the-art code LLMs to code transformations vary significantly across tasks and transformation operators. In addition, the LLMs exhibit better robustness to the semantic preserving transformations than their sensitivity to the semantic non-preserving transformations. These results highlight a need to enhance the model's capabilities of understanding code semantics, especially sensitivity.

\printcredits

\bibliographystyle{elsarticle-num}

\bibliography{references}

\end{document}